\pdfoutput=1 
\RequirePackage{ifpdf}
\ifpdf 
\documentclass[pdftex]{sigma}
\else
\documentclass{sigma}
\fi

\numberwithin{equation}{section} \numberwithin{theorem}{section}
\numberwithin{proposition}{section} \numberwithin{lemma}{section}
\numberwithin{definition}{section}

\begin{document}

\allowdisplaybreaks

\renewcommand{\thefootnote}{$\star$}

\renewcommand{\PaperNumber}{013}

\FirstPageHeading

\ShortArticleName{Derivations of the Moyal Algebra and
Noncommutative Gauge Theories}

\ArticleName{Derivations of the Moyal Algebra\\ and Noncommutative
Gauge Theories\footnote{This paper is a contribution to the
Proceedings of the XVIIth International Colloquium on Integrable
Systems and Quantum Symmetries (June 19--22, 2008, Prague, Czech
Republic). The full collection is available at
\href{http://www.emis.de/journals/SIGMA/ISQS2008.html}{http://www.emis.de/journals/SIGMA/ISQS2008.html}}}

\Author{Jean-Christophe WALLET}

\AuthorNameForHeading{J.-C. Wallet}

\Address{Laboratoire de Physique Th\'eorique, B\^at.\ 210, CNRS, Universit\'e Paris-Sud 11,\\
F-91405 Orsay Cedex, France}
\Email{\href{mailto:jean-christophe.wallet@th.u-psud.fr}{jean-christophe.wallet@th.u-psud.fr}}

\ArticleDates{Received October 29, 2008, in f\/inal form January
17, 2009; Published online January 30, 2009}

\Abstract{The dif\/ferential calculus based on the derivations of
an associative algebra underlies most of the noncommutative
f\/ield theories considered so far. We review the essential
properties of this framework and the main features of
noncommutative connections in the case of non graded associative
unital algebras with involution. We extend this framework to the
case of ${\mathbb{Z}}_2$-graded unital involutive algebras. We
show, in the case of the Moyal algebra or some related
${\mathbb{Z}}_2$-graded version of it, that the derivation based
dif\/ferential calculus is a suitable framework to construct
Yang--Mills--Higgs type models on Moyal (or related) algebras, the
covariant coordinates having in particular a natural
interpretation as Higgs f\/ields. We also exhibit, in one
situation, a link between the renormalisable NC $\varphi^4$-model
with harmonic term and a gauge theory model. Some possible
consequences of this are brief\/ly discussed.}

\Keywords{noncommutative geometry; noncommutative gauge theories}

\Classification{81T75; 81T13}

\renewcommand{\thefootnote}{\arabic{footnote}}
\setcounter{footnote}{0}

\section{Introduction}
A class of noncommutative (NC)  f\/ield theories
\cite{Douglas:2001ba, Szabo} (for general reviews on
noncommutative geo\-met\-ry, see \cite{CONNES, CM}) came under
increasing scrutiny after 1998 when it was argued
\cite{Schomerus,Seiberg:1999vs} that string theory might have
ef\/fective regimes related to noncommutative f\/ield theories
(NCFT) def\/ined on a NC version of f\/lat  4-dimensional space.
This latter is the Moyal space (see
e.g.~\mbox{\cite{Gracia-Bondia:1987kw,Varilly:1988jk}}) which has
constant commutators between space coordinates. It was further
noticed~\cite{Minwalla:1999px,Chepelev:1999tt} that the simplest
NC $\varphi^4$ model, ($\varphi$ real-valued) on the 4-dimensional
Moyal space is not renormalisable due to the Ultraviolet/Infrared
(UV/IR) mixing
\cite{Minwalla:1999px,Chepelev:1999tt,Matusis:2000jf}. This
phenomenon stems from the existence of nonplanar diagrams that are
UV f\/inite but nevertheless develop IR singularities which when
inserted into higher order diagrams are not of renormalisable
type~\cite{Douglas:2001ba,Szabo}. A~solution to this problem was
proposed in 2004 \cite{Grosse:2004yu,Grosse:2003aj}. It amounts to
supplement the initial action with a simple harmonic oscillator
term leading to a fully renormalisable NCFT (for recent reviews,
see e.g.~\cite{vince,Wallet:2007 em}). This result seems to be
related to the covariance of the model under the so called
Langmann--Szabo duality~\cite{Langmann:2002cc}. Other
renormalisable NC matter f\/ield theories have then been
identif\/ied  \cite{Grosse:2003nw,Langmann:2003if,
Langmann:2003cg, Vignes-Tourneret:2006nb} and studies of the
properties of their renormalisation group f\/lows have been
carried out~\cite{beta1, Lakhoua:2007ra, beta}.

So far, the construction of a renormalisable gauge theory on 4-$D$
Moyal spaces remains a~problem. The naive NC version of the
Yang--Mills action has UV/IR mixing, which stems from the
occurrence of an IR singularity in the polarisation tensor. From a
standard one-loop calculation, we easily infer that
\begin{equation}
\omega_{\mu\nu}(p)\sim
(D-2)\Gamma\left({{D}\over{2}}\right){{{\tilde{p}}_\mu{\tilde{p}}_\nu}\over{{\pi^{D/2}(\tilde{p}}^2)^{D/2}}}+\cdots
,\qquad p\to0, \label{eq:singul1}
\end{equation}
where ${\tilde{p}}_\mu$$\equiv$$\Theta_{\mu\nu}p_\nu$,
$\Theta_{\mu\nu}$ is the symplectic matrix of the 4-$D$ Moyal
algebra and $\Gamma(z)$ denotes the Euler function. This
singularity, albeit transverse in the sense of the
Slavnov--Taylor--Ward identities, does not correspond to some
gauge invariant term. This implies that the recent alternative
solution to the UV/IR mixing proposed for the NC $\varphi^4$ model
in \cite{GMRT}, cannot be extended straighforwardly (if possible
at all) to the gauge theories. Note that an attempt to reach this
goal has been proposed in \cite{gieres}. It amounts to add to the
naive Yang--Mills action a counterterm which remains invariant
under a def\/ined BRST symmetry but however can be viewed as a
formal inf\/inite series in the gauge potential $A_\mu$. The
actual ef\/fect of this counterterm on the UV/IR mixing of the
modif\/ied gauge action is still unclear and deserves further
investigations.

Recently, an extension of the harmonic solution to the case of
gauge theories has been proposed in \cite{de Goursac:2007gq} and
\cite{Grosse:2007qx} (for various reviews, see
\cite{deGoursac:2007qi,Wallet:2007 em,Wohl2007}). These works have
singled out, as potential candidate for renormalisable gauge
theory on 4-$D$ Moyal space, the following generic action
\begin{equation}
S=\int d^4x \left(\frac{1}{4}F_{\mu\nu}\star F_{\mu\nu}
+\frac{\Omega^2}{4}\{{\cal{A}}_\mu,{\cal{A}}_\nu\}^2_\star
+{\kappa}{\cal{A}}_\mu\star{\cal{A}}_\mu\right),
\label{eq:decadix1}
\end{equation}
where $F_{\mu\nu}=\partial_\mu A_\nu-\partial_\nu
A_\mu-i[A_\mu,A_\nu]_\star$, $A_\mu$ is the gauge potential,
$\star$ denotes the associative product on the Moyal algebra,
$[a,b]_\star\equiv a\star b-b\star a$, $\{a,b\}_\star\equiv a\star
b+b\star a$ and ${\cal{A}}_\mu$ is the covariant coordinates, a
natural gauge covariant tensorial form stemming from the existence
of a canonical gauge invariant connection in the present NC
framework, as we will recall in a~while (see \cite{de
Goursac:2007gq, Wallet:2007 em}). The 2nd and 3rd terms
in~\eqref{eq:decadix1} may be viewed as ``gauge counterparts'' of
the harmonic term of \cite{Grosse:2004yu}. This action has
interesting properties \cite{de Goursac:2007gq,Grosse:2007qx}
deserving further studies. For instance, gauge invariant mass
terms for the gauge f\/ields are allowed even in the absence of
Higgs mechanism. Besides, the covariant coordinates appear to bear
some similarity with Higgs f\/ields. This feature will be examined
more closely in the course of the discussion. Unfortunately, the
action~\eqref{eq:decadix1} has a non-trivial vacuum which
complicates the study of its actual
renormalisability~\cite{vacuumym:2008}. Notice that non trivial
vacuum conf\/igurations also occur within NC scalar models with
harmonic term as shown in~\cite{deGoursac:2007uv}.

The relevant algebraic framework describing (most of) the
classical features of the NC actions considered so far is provided
by the dif\/ferential calculus based on the derivations. This
framework has been established in \cite{DBV-1, DBV-2,
Dubois-Violette:1989vr} and \cite{DBV-3}. For an exhaustive
review, see \cite{DBV-4} and references therein. The derivation
based dif\/ferential calculus underlies the f\/irst prototypes of
NC (matrix-valued) f\/ield theories \cite{Dubois-Violette:1989vq,
DBV-M,Masson:1999}. For a review, see \cite{Masson:Orsay}. As far
as the Moyal spaces are concerned, the ``minimal''
derivation-based dif\/ferential calculus  generated by the
``spatial derivations'' $\partial_\mu$ underlies (most of) the
works that appeared in the literature, so far. This dif\/ferential
calculus is not unique but can be modif\/ied in numerous ways.
Among these, a simple modif\/ication of the minimal dif\/ferential
calculus has been shown recently to give rise to interesting
features~\cite{can-mas-wal}.

The purpose of this paper is to show that the use of the
dif\/ferential calculus based on the derivations of the Moyal
algebra or some related ${\mathbb{Z}}_2$-graded version of it
permits one to exhibit interesting features related to gauge
theories def\/ined on Moyal space. We show, at least within two
non trivial examples, that the derivation based dif\/ferential
calculus is a natural framework to construct Yang--Mills--Higgs
type models on Moyal (or related) algebras, the so-called
``covariant coordinates'' \cite{Douglas:2001ba,Szabo} having a
natural interpretation as Higgs f\/ields. We also exhibit, in one
situation, a link between the renormalisable NC $\varphi^4$-model
with harmonic term \cite{Grosse:2004yu,Grosse:2003aj} and a~gauge
theory model built from the square of a curvature. Some possible
consequences of this are brief\/ly discussed.

The paper is organised as follows. In Section~\ref{section2.1} of
this paper, we review brief\/ly the main properties of the
dif\/ferential calculus based on the derivations of an associative
unital algebra and introduce a def\/inition of a NC connection on
a module over the algebra, as a natural generalisation of ordinary
connections. The specif\/ic properties and simplif\/ications
occurring when the module is equal to the algebra, which is the
case relevant for the NCFT, are detailed in
Section~\ref{section2.2}. In Section~\ref{gradedcalc-section}, we
extend the analysis to ${\mathbb{Z}}_2$-graded algebras. We
construct a~${\mathbb{Z}}_2$-graded dif\/ferential calculus based
on the graded derivations of the algebra, set a def\/inition of NC
connection, the corresponding curvature and (unitary) gauge
transformations. In Section~\ref{section3}, we focus on the Moyal
algebra ${\cal{M}}$. We consider the dif\/ferential calculus based
on the maximal subalgebra of the derivations of ${\cal{M}}$ whose
elements can be related to (inf\/initesimal) symplectomorphisms.
Then, a direct application of the results of
Section~\ref{section2.2} leads to a natural construction of
Yang--Mills--Higgs models def\/ined on ${\cal{M}}$ where the
``covariant coordinates'' used in the physics literature can be
naturally interpreted as Higgs f\/ields, due to the existence of
a~gauge invariant canonical connection. In Section~\ref{section4}
we compare in detail the salient mathema\-ti\-cal features
underlying the NC dif\/ferential calculus of
Section~\ref{section3} to those for the NC geometry stemming from
the f\/inite dimensional matrix algebra $M_n({\mathbb{C}})$ as
well as for the algebra of matrix valued functions
$C^\infty(M)\otimes M_n({\mathbb{C}})$. In some sense, the case
considered in Section~\ref{section3} interpolates between these
two latter situations. The classical properties of the NC
Yang--Mills--Higgs actions are also analysed. Explicit one-loop
computation of the vacuum polarisation tensor shows that this
latter still exhibits an IR singularity of the type given in
\eqref{eq:singul1}. In Section~\ref{section5} we consider
a~${\mathbb{Z}}_2$-graded version of the Moyal algebra ${\cal{M}}$
built from two copies of ${\cal{M}}$ and apply the gene\-ral
results derived in Section~\ref{gradedcalc-section} to the
construction of a~dif\/ferential calculus generated by
a~${\mathbb{Z}}_2$-graded extension of the derivation algebra
considered in~Section~\ref{section3}. We show that the gauge
theory action built from the ``square'' of the resulting curvature
involves as contributions both the action~\eqref{eq:decadix1}
derived in~\cite{de Goursac:2007gq} and \cite{Grosse:2007qx} as
well the renormalisable NC $\varphi^4$-model with harmonic term
elaborated in \cite{Grosse:2004yu,Grosse:2003aj}, therefore
exhibiting a link between this latter renormalisable NC scalar
theory and gauge theories. Finally, a summary of all the main
results is presented in Section~\ref{section6}.

\section[Differential calculus based on derivations]{Dif\/ferential calculus based on derivations}\label{section2}

\subsection{General properties}\label{section2.1}

Let ${\mathbb{A}}$ be an associative $*$-algebra with unit
${\mathbb{I}}$ and center ${\cal{Z}}({\mathbb{A}})$. We denote the
involution by $a\mapsto a^\dag$, $\forall\, a\in {\mathbb{A}}$.
The dif\/ferential calculus based on the derivations of
${\mathbb{A}}$ is a natural NC generalisation of the usual de Rham
dif\/ferential calculus on a manifold. Basically, the role of the
vector f\/ields is now played by the derivations of the algebra.
In this subsection, we collect the main properties that will be
used in this paper. More details can be found in \cite{DBV-1,
Dubois-Violette:1989vr, DBV-2, DBV-3}.

\begin{definition}\label{def1}
The vector space of derivations of ${\mathbb{A}}$ is the space of
linear maps def\/ined by ${\text{\textup{Der}}}({\mathbb{A}}) = \{
X : {\mathbb{A}} \rightarrow {\mathbb{A}} \, / \, X(ab) = X(a) b +
a X(b), \; \forall\, a,b\in {\mathbb{A}}\}$. The derivation
$X\in{\text{\textup{Der}}}({\mathbb{A}})$ is called real if
$(X(a))^\dag=X(a^\dag)$, $\forall\, a\in{\mathbb{A}}$.
\end{definition}

The essential properties of the spaces of derivations of
${\mathbb{A}}$ can be summarised in the following proposition.
\begin{proposition}\label{prop1}
${\text{\textup{Der}}}({\mathbb{A}})$ is a
${\cal{Z}}({\mathbb{A}})$-module for the product $(fX )a = f(X
a)$, $\forall \, f \in {\cal{Z}}({\mathbb{A}})$, $\forall\,  X \in
{\text{\textup{Der}}}({\mathbb{A}})$ and a Lie algebra for the
bracket $[X, Y ]a = X  Y a - Y X a$, $\forall\, X,Y \in
{\text{\textup{Der}}}({\mathbb{A}})$. The vector subspace of inner
derivations is defined by ${\textup{Int}}({\mathbb{A}}) = \{
{\textup{Ad}}_a : b \mapsto [a,b]\, / \, a \in {\mathbb{A}}\}
\subset {\text{\textup{Der}}}({\mathbb{A}})$. It is a
${\cal{Z}}({\mathbb{A}})$-submodule and Lie ideal. The vector
subspace of outer derivations is
${\textup{Out}}({\mathbb{A}})={\text{\textup{Der}}}({\mathbb{A}})/{\textup{Int}}({\mathbb{A}})$.
The following canonical short exact sequence of Lie algebras and
${\cal{Z}}({\mathbb{A}})$-modules holds:
$0\longrightarrow{\textup{Int}}({\mathbb{A}})\longrightarrow{\text{\textup{Der}}}
({\mathbb{A}})\longrightarrow{\textup{Out}}({\mathbb{A}})\longrightarrow0$.
\end{proposition}
The main features of the dif\/ferential calculus based on
${\text{\textup{Der}}}({\mathbb{A}})$ are involved in the
following proposition. Notice that both the Lie algebra structure
and the ${\cal{Z}}({\mathbb{A}})$-module structures for
${\text{\textup{Der}}}({\mathbb{A}})$ are used as essential
ingredients in the construction.

\begin{proposition} \label{prop2}
Let $\underline{\Omega}^n_{{\text{\textup{Der}}}}({\mathbb{A}})$
denote the space of ${\cal{Z}}({\mathbb{A}})$-multilinear
antisymmetric maps from ${\text{\textup{Der}}}({\mathbb{A}})^n$ to
${\mathbb{A}}$, with
$\underline{\Omega}^0_{{\text{\textup{Der}}}}({\mathbb{A}}) =
{\mathbb{A}}$ and let
$\underline{\Omega}^\bullet_{{\text{\textup{Der}}}}({\mathbb{A}})
=\textstyle \bigoplus_{n \geq 0}
\underline{\Omega}^n_{{\text{\textup{Der}}}}({\mathbb{A}})$. Then
($\underline{\Omega}^\bullet_{{\text{\textup{Der}}}}({\mathbb{A}})$,
$\times$, ${\hat{d}}$) is a ${\mathbb{N}}$-graded differential
algebra with the product $\times$ on
$\underline{\Omega}^\bullet_{{\text{\textup{Der}}}}({\mathbb{A}})$
and differential
${\hat{d}}:\underline{\Omega}^n_{{\text{\textup{Der}}}}({\mathbb{A}})\to\underline{\Omega}^{n+1}_{{\text{\textup{Der}}}}({\mathbb{A}})$
satisfying ${\hat{d}}^2=0$, respectively defined for
$\forall\omega\in\underline{\Omega}^p_{{\text{\textup{Der}}}}({\mathbb{A}}),\
\eta\in\underline{\Omega}^q_{{\text{\textup{Der}}}}({\mathbb{A}})$
by:
\begin{gather}
(\omega\times\eta)(X_1,\dots, X_{p+q}) =
 \frac{1}{p!q!} \sum_{\sigma\in {\mathfrak{S}}_{p+q}}\!\! (-1)^{{\textup{sign}}(\sigma)}
\omega(X_{\sigma(1)},\dots, X_{\sigma(p)})
\eta(X_{\sigma(p+1)},\dots,
X_{\sigma(p+q)}), \nonumber
\\
{\hat{d}}\omega(X_1,\dots, X_{p+1}) = \sum_{i=1}^{p+1} (-1)^{i+1}
X_i \omega( X_1,\dots \vee_i \dots, X_{p+1}) \nonumber
\\
\phantom{{\hat{d}}\omega(X_1,\dots, X_{p+1}) =}{}  + \sum_{1\leq i
< j \leq p+1} (-1)^{i+j} \omega( [X_i, X_j],\dots \vee_i \dots
\vee_j \dots, X_{p+1}), \label{eq:koszul}
\end{gather}
where $\sigma$ denotes permutation and the symbol $\vee_i$
indicates that $X_i$ is omitted.
\end{proposition}
It turns out that a dif\/ferential calculus can also be built from
suitable subalgebras of ${\text{\textup{Der}}}({\mathbb{A}})$. The
following proposition holds \cite{DBV-1, DBV-2}

\begin{proposition}\label{prop3}
Let ${\cal{G}} \subset {\text{\textup{Der}}}({\mathbb{A}})$ denote a Lie subalgebra which is also a ${\cal{Z}}({\mathbb{A}})$-submodule. Then, a restricted derivation-based differential calculus $\underline{\Omega}^\bullet_{\cal{G}}({\mathbb{A}})$ can be built from ${\cal{G}}$. It is obtained from Proposition~{\rm \ref{prop2}} by replacing the set of $\underline{\Omega}^n_{\text{\textup{Der}}}({\mathbb{A}})$, $\forall \, n\in{\mathbb{N}}$ by the set of ${\cal{Z}}({\mathbb{A}})$-multilinear antisymmetric maps from ${\cal{G}}^n$ to ${\mathbb{A}}$ for $n \geq 0$ and still using 
\eqref{eq:koszul}.
\end{proposition}
From now on, we will not write explicitly the product of forms
$\times$. It should be obvious from the context when the relevant
type of product is used.

In this paper, we will consider a natural NC generalisation of
ordinary connections, as introduced in \cite{DBV-1,
Dubois-Violette:1989vr, DBV-2} to which we refer for more details.
It uses left or right f\/inite projective modules on the
associative algebra. Notice that alternative NC extensions of
connections based on bimodules were considered in~\cite{DBV-3}.
From now on, we denote by ${\mathbb{M}}$ a right
${\mathbb{A}}$-module. Let
$h:{\mathbb{M}}\otimes{\mathbb{M}}\to{\mathbb{A}}$ denote a
Hermitian structure{\footnote{Recall that a Hermitian structure is
a sesquilinear map such that $h(m_1,m_2)^\dag=h(m_2,m_1)$,
$h(ma_1,ma_2)=a_1^\dag h(m1,m2)a_2$, $\forall \,
m_1,m_2\in{\mathbb{M}}$, $\forall\, a_1,a_2\in{\mathbb{A}}$.}} on
${\mathbb{A}}$. The connection, curvature and gauge
transformations are given as follows:
\begin{definition} \label{def2}
A NC connection on ${\mathbb{M}}$ is a linear map ${\nabla} :
{\mathbb{M}}\times {\textup{Der}}(\mathbb{A}) \rightarrow
{\mathbb{M}}$ satisfying:
\begin{gather}
{\nabla}_X (m a) = mX( a) + {\nabla}_X (m) a,\qquad
{\nabla}_{fX}( m) = f{\nabla}_X (m),\nonumber\\
{\nabla}_{X + Y} (m) = {\nabla}_X (m) + {\nabla}_Y (m),
\label{eq:connect-leib}
\end{gather}
$\forall\, X,Y \in {\text{\textup{Der}}}({\mathbb{A}})$,
$\forall\, a \in {\mathbb{A}}$, $\forall \, m \in {\mathbb{M}}$,
$\forall\, f \in {\cal{Z}}({\mathbb{A}})$.  A Hermitian NC
connection is a NC connection satisfying in addition
$X(h(m_1,m_2))=h(\nabla_X(m_1),m_2)+h(m_1,\nabla_X(m_2))$,
$\forall \, m_1,m_2\in{\mathbb{M}}$, and for any real derivation
in ${\text{\textup{Der}}}({\mathbb{A}})$. The curvature of
${\nabla}$ is the linear map ${F}(X, Y) : {\mathbb{M}} \rightarrow
{\mathbb{M}}$ def\/ined~by
\begin{gather*}
 {F}(X, Y) m = [ {\nabla}_X,{\nabla}_Y ] m - {\nabla}_{[X, Y]}m,\qquad \forall \, X, Y \in {\text{\textup{Der}}}({\mathbb{A}}).
\end{gather*}
\end{definition}

\begin{definition}\label{def3}
The gauge group of ${\mathbb{M}}$ is def\/ined as the group of
automorphisms of ${\mathbb{M}}$ as a~right ${\mathbb{A}}$-module.
\end{definition}

\begin{proposition}\label{prop4}
For any $g$ in the gauge group of ${\mathbb{M}}$ and for any NC
connection ${\nabla}$, the map ${\nabla}^g_X = g^{-1}\circ
{\nabla}_X \circ g : {\mathbb{M}} \rightarrow {\mathbb{M}}$
defines a NC connection. Then, one has ${F}(X,Y)^g=g^{-1}\circ
{F}(X,Y) \circ g$.
\end{proposition}

It is convenient to require that the gauge transformations are
compatible with the Hermitian structure, that is $h(g(m_1),
g(m_2))=h(m_1,m_2)$. This def\/ines a NC analogue of unitary gauge
transformations. From now on, we will only consider unitary gauge
transformations.

\subsection{The free module case}\label{section2.2}

In the special case where ${\mathbb{M}}={\mathbb{A}}$, that will
be the case of interest for the ensuing discussion, additional
simplif\/ications occur. It is further convenient to choose the
canonical Hermitian structure $h_0(a_1,a_2)=a_1^\dag a_2$.

\begin{proposition}\label{prop5}
Assume that ${\mathbb{M}}={\mathbb{A}}$ and
$h_0(a_1,a_2)=h(a_1,a_2)=a_1^\dag a_2$. Then:
\begin{enumerate}\itemsep=0pt
\item[$i)$] Any NC connection is entirely determined by
$\nabla_X({\mathbb{I}})$ via $\nabla_X(a)=
\nabla_X({\mathbb{I}})a+X(a)$, $\forall \,
X\in{\text{\textup{Der}}}({\mathbb{A}})$, $\forall\,
a\in{\mathbb{A}}$. The $1$-form connection
$A\in\underline{\Omega}^1_{\text{\textup{Der}}}({\mathbb{A}})$ is
defined by $A:X\to A(X)=\nabla_X({\mathbb{I}})$, $\forall\,
X\in{\text{\textup{Der}}}({\mathbb{A}})$. \item[$ii)$] Assume that
the derivations $X$ are real. Then, a NC connection is Hermitian
when $\nabla_X({\mathbb{I}})^\dag=-\nabla_X({\mathbb{I}})$.
\item[$iii)$] The gauge group can be identified with the group of
unitary elements of ${\mathbb{A}}$, ${\cal{U}}({\mathbb{A}})$ and
one has
$\nabla_X({\mathbb{I}})^g=g^\dag\nabla_X({\mathbb{I}})g+g^\dag
X(g)$, ${F}(X,Y)^g=g^\dag{F}(X,Y)g$, $\forall\, X,
Y\in{\text{\textup{Der}}}({\mathbb{A}})$, $\forall\,
a\in{\mathbb{A}}$.
\end{enumerate}
\end{proposition}
\begin{proof}
$i)$ follows directly from Def\/inition \ref{def2} (set
$m={\mathbb{I}}$ in the 1st of~\eqref{eq:connect-leib}). Note that
the relation $\nabla_X(a)= \nabla_X({\mathbb{I}})a+X(a)$ can be
reobtained from the map
$\nabla:\underline{\Omega}^0_{\text{\textup{Der}}}({\mathbb{A}})\to\underline{\Omega}^1_{\text{\textup{Der}}}({\mathbb{A}})$,
$\nabla(a)=Aa+{\hat{d}}a$ with $A:X\to
A(X)=\nabla_X({\mathbb{I}})$. For $ii)$, one has
$(\nabla_X(a_1)^\dag a_2+a_1^\dag\nabla_X(a_2)$$=$$X(a_1^\dag
a_2)+a_1^\dag(\nabla_X({\mathbb{I}})^\dag+\nabla_X({\mathbb{I}}))a_2$
where the last equality stems from the expression for
$\nabla_X(a)$ given in $i)$ and the fact that $X$ is assumed to be
real. From this follows ii). For $iii)$, use
Def\/inition~\ref{def3} and compatibility of gauge transformations
with $h_0$ which gives $g(a)=g({\mathbb{I}})a$ and
$h_0(g(a_1),g(a_2))=a_1^\dag g({\mathbb{I}})^\dag
g({\mathbb{I}})a_2=h_0(a_1,a_2)$. Then, the gauge transformations
for $\nabla_X({\mathbb{I}})$ and the curvature stems from
Proposition~\ref{prop4}, the expression for $\nabla_X(a)$ in $i)$
and the expression for ${F}(X,Y)$.
\end{proof}

\begin{definition}\label{tensorform}
A tensor 1-form is a 1-form having the following gauge
transformations:
\begin{equation*}
 {\cal{A}}^g=g^\dag{\cal{A}}g,\qquad \forall\, g\in{\cal{U}}({\mathbb{A}}).
\end{equation*}
\end{definition}
There is a special situation where canonical gauge invariant
connections can show up, as indicated in the following
proposition.

\begin{proposition}\label{prop2.9}
Assume that there exists
$\eta\in\underline{\Omega}^1_{\text{\textup{Der}}}({\mathbb{A}})$,
such that ${\hat{d}}a=[\eta,a]$, $\forall \, a\in{\mathbb{A}}$.
Consider the map $\nabla^{\rm
inv}:\underline{\Omega}^0_{\text{\textup{Der}}}({\mathbb{A}})\to\underline{\Omega}^1_{\text{\textup{Der}}}({\mathbb{A}})$,
$\nabla^{\rm inv}(a)={\hat{d}}a-\eta a$, $\forall \,
a\in{\mathbb{A}}$, so that $\nabla^{\rm inv}_X(a)=X(a)-\eta(X) a$.
Then, the following properties hold:
\begin{enumerate}\itemsep=0pt
\item[$i)$] $\nabla^{\rm inv}$ defines a connection which is gauge
invariant, called the canonical connection.

\item[$ii)$] For any NC connection $\nabla$,
${\cal{A}}\equiv\nabla-\nabla^{\rm inv}=A+\eta$ defines a tensor
form. ${\cal{A}}(X)$, $\forall \,
X\in{\text{\textup{Der}}}({\mathbb{A}})$ are called the covariant
coordinates of $\nabla$.
\end{enumerate}
\end{proposition}

\begin{proof}
Since any 1-form can serve as def\/ining a connection in view of
Proposition~\ref{prop5}, $\nabla^{\rm inv}(a)={\hat{d}}a-\eta a$
is a connection. Notice that it reduces to $\nabla^{\rm
inv}(a)=-a\eta$, since ${\hat{d}}a=[\eta,a]$. Then, one has
$(\nabla^{\rm inv})^g(a)=g^\dag\nabla^{\rm
inv}(ga)=g^\dag(d(ga)-\eta ga)$
$=g^\dag(-ga\eta)=-a\eta=\nabla^{\rm inv}(a)$, which shows~$i)$.
The property~$ii)$ stems simply from Def\/inition~\ref{tensorform}
and the gauge transformations of a~NC connection.
\end{proof}

The existence of canonical connections translates into some rather
general properties of the curvatures, in particular the curvature
for the canonical connection. Gauge theories def\/ined on Moyal
spaces are a particular example of this, as shown in the next
section.
\begin{lemma} \label{lemma}
Let ${F}^{\rm inv}{(X,Y)}\equiv\eta{[X,Y]}-[\eta(X),\eta(Y)]$
denote the curvature for the canonical connection. Assume again
that there exists
$\eta\in\underline{\Omega}^1_{\text{\textup{Der}}}({\mathbb{A}})$,
such that ${\hat{d}}a=[\eta,a]$, $\forall\, a\in{\mathbb{A}}$.
Then, the following properties hold:
\begin{enumerate}\itemsep=0pt
\item[$i)$] ${F}^{\rm
inv}{(X,Y)}\equiv\eta{[X,Y]}-[\eta(X),\eta(Y)]\in{\cal{Z}}({\mathbb{A}})$.
\item[$ii)$] The curvature of any NC connection defined by the
tensor $1$-form ${\cal{A}}$ can be written as
\begin{equation*}
{F}{(X,Y)}=([{\cal{A}}(X),{\cal{A}}(Y)]-{\cal{A}}{[X,Y]})-([\eta(X),\eta(Y)]-\eta{[X,Y]}),\  \ \ \forall\, X,Y\in{\text{\textup{Der}}}({\mathbb{A}}). 
\end{equation*}
\end{enumerate}
\end{lemma}

\begin{proof}
First, from the def\/inition of $\nabla^{\rm inv}(a)$ in
Proposition~\ref{prop2.9}, one infers that the 2-form curvature
associated to the canonical connection is $F^{\rm
inv}(a)\equiv\nabla^{\rm inv}(\nabla^{\rm
inv}(a))=-({\hat{d}}\eta-\eta\eta)(a)$, $\forall\,
a\in{\mathbb{A}}$. Then, one obtains $F^{\rm
inv}(X,Y)=\eta{[X,Y]}-[\eta(X),\eta(Y)]$. Then, from
${\hat{d}}a=[\eta,a]$ and ${\hat{d}}^2=0$, one has
${\hat{d}}({\hat{d}}a)\!=\!{\hat{d}}(\eta
a-a\eta)\!=\![{\hat{d}}\eta,a]-[\eta,{\hat{d}}a]\!=\![{\hat{d}}\eta,a]-[\eta,[\eta,a]]\!=\![{\hat{d}}\eta-\eta\eta,a]$.
From this follows the property~$i)$. Next, one has
$\nabla_X(a)={\cal{A}}(X)a-a\eta(X)$ so that
$[\nabla_X,\nabla_Y](a)=[{\cal{A}}(X),{\cal{A}}(Y)]a-a[\eta(X),\eta(Y)]$.
Therefore
${F}{(X,Y)}(a){=}([{\cal{A}}(X),{\cal{A}}(Y)]-{\cal{A}}({[X,Y])})a-a([\eta(X),\eta(Y)]-\eta({[X,Y]}))$.
This last expression, combined with the property~$i)$
implies~$ii)$.
\end{proof}

Notice that this proposition will be relevant when considering the
NC gauge theories on Moyal algebras. A somewhat similar (but not
identical!) situation occurs for the algebra of
matrices~$M_n({\mathbb{C}})$ as well as for the algebra of matrix
valued functions.

\subsection[Graded differential calculus]{Graded dif\/ferential calculus} \label{gradedcalc-section}

In this subsection, we extend the previous algebraic scheme to the
case of dif\/ferential calculus based on the derivations of a
graded algebra. We consider only the ${\mathbb{Z}}_2$-graded case.
The extension to graduations based on additive groups as well as
to more general structures \cite{scheunert} will be reported in a
separate work \cite{ax-t-jc}.

To f\/ix the notations, let ${\mathbb{A}}^\bullet$ be a
${\mathbb{Z}}_2$-graded associative unital $*$-algebra, namely
${\mathbb{A}}^\bullet=\bigoplus_{\alpha=\ 0,1}{\mathbb{A}}^\alpha$
where ${\mathbb{A}}^\alpha=\{ a\in{\mathbb{A}}^\bullet\, /\,
|a|=\alpha\}$, $|a|$ denoting the homogeneous degree of $a$, and
${\mathbb{A}}^\alpha{\mathbb{A}}^\beta\subseteq{\mathbb{A}}^{(\alpha+\beta)({\rm
mod}\, 2)}$. The graded bracket on ${\mathbb{A}}^\bullet$ is
$[a,b]_\bullet\equiv ab-(-1)^{|a||b|}b a$, $\forall\,
a,b\in{\mathbb{A}}^\bullet$. Let ${\cal{Z}}({\mathbb{A}}^\bullet)$
denote the center of ${\mathbb{A}}^\bullet$, a graded-commutative
algebra ($a\in{\cal{Z}}({\mathbb{A}}^\bullet)\iff[a,b]_\bullet=0$,
$\forall\, b\in{\mathbb{A}}^\bullet$). Graded modules will be
needed in the present framework. Recall that they are graded
vector spaces
${\mathbb{M}}^\bullet=\bigoplus_{\alpha=0,1}{\mathbb{M}}^\alpha$
satisfying
${\mathbb{M}}^\alpha{\mathbb{A}}^\beta\subseteq{\mathbb{M}}^{\alpha+\beta}$
and
${{\mathbb{A}}^\beta\mathbb{M}}^\alpha\subseteq{\mathbb{M}}^{\alpha+\beta}$
respectively for a~right- and left-${\mathbb{A}}^\bullet$-module.

\begin{definition}\label{def4}
The ${\mathbb{Z}}_2$-graded vector space of derivations of
${\mathbb{A}}^\bullet$ is
${\text{\textup{Der}}}({\mathbb{A}}^\bullet)=\bigoplus_{\alpha=
0,1}{\text{\textup{Der}}}^\alpha({\mathbb{A}}^\bullet)$ where
${\text{\textup{Der}}}^\alpha({\mathbb{A}}^\bullet)$ is the vector
space of linear maps $X$ of homogeneous degree $|X|=\alpha$
def\/ined by
${\text{\textup{Der}}}^\alpha({\mathbb{A}}^\bullet)=\{
X:{\mathbb{A}}^\beta \to {\mathbb{A}}^{\alpha+\beta}\, /\,
X(ab)=X(a)b+(-1)^{\alpha\beta}aX(b), \forall\,
a\in{\mathbb{A}}^\beta, \forall\, b\in{\mathbb{A}}^\bullet\}$.
\end{definition}

Most of the properties of the spaces of derivations in the
non-graded case can be extended to the graded situation. In the
rest of this paper, we will consider the case where the graded
module is the graded algebra itself, i.e.\
${\mathbb{M}}^{\bullet}={\mathbb{A}}^\bullet$.

\begin{proposition}\label{prop7}
${\text{\textup{Der}}}({\mathbb{A}}^\bullet)$ is a graded right
${\cal{Z}}({\mathbb{A}}^\bullet)$-module for the product
$(zX)(a)=z(X(a))$, a graded left
${\cal{Z}}({\mathbb{A}}^\bullet)$-module for the product
$(Xz)(a)=(-1)^{|a||z|}(X(a))z$, $\forall\, z\in
{\cal{Z}}({\mathbb{A}}^\bullet)$, $\forall\, X\in
{\text{\textup{Der}}}({\mathbb{A}}^\bullet)$ of homogeneous degree
and a ${\mathbb{Z}}_2$-graded Lie algebra of the graded bracket
$[X,Y]_\bullet=XY-(-1)^{|X||Y|}YX$, $\forall\,
X,Y\in{\text{\textup{Der}}}({\mathbb{A}}^\bullet)$. The
${\mathbb{Z}}_2$-graded vector subspaces of inner derivations is
defined by ${\textup{Int}}^\bullet({\mathbb{A}}) = \{
{\textup{Ad}}_a : b \mapsto [a,b]_\bullet\, / \, a \in
{\mathbb{A}}^\bullet\} \subset
{\text{\textup{Der}}}^\bullet({\mathbb{A}})$. It is a
${\mathbb{Z}}_2$-graded
${\cal{Z}}({\mathbb{A}}^\bullet)$-submodules and Lie subalgebra.
The ${\mathbb{Z}}_2$-graded vector subspace of outer derivations
is defined as
${\textup{Out}}({\mathbb{A}}^\bullet)={\text{\textup{Der}}}({\mathbb{A}}^\bullet)/{\textup{Int}}({\mathbb{A}}^\bullet)$.
\end{proposition}

\begin{proof}
The properties follow from straightforward calculations.
\end{proof}

\begin{definition}\label{def5}
The space
$\Omega_{\text{\textup{Der}}}^{n,\alpha}({\mathbb{A}}^\bullet)$,
$\alpha\in\{0,1\}$, $n\in{\mathbb{N}}$ denotes the vector space of
$n$-linear maps $\omega$ with homogenous degree $\alpha$,
$\omega:({\text{\textup{Der}}}({\mathbb{A}}^\bullet))^{\times
n}\to{\mathbb{A}}^\bullet$ def\/ined by
\begin{enumerate}
\itemsep=0pt \item[$i)$]
$\Omega_{\text{\textup{Der}}}^{0,\alpha}({\mathbb{A}}^\bullet)={\mathbb{A}}^\alpha$,
$\forall\,\alpha\in\{0,1\}$ and,
$\forall\,\omega\in\Omega_{\text{\textup{Der}}}^{n,\alpha}({\mathbb{A}}^\bullet)$,
$\forall \, X_i\in{\text{\textup{Der}}}({\mathbb{A}}^\bullet)$ of
homogeneous degree,

\item[$ii)$] $\omega(X_1,\dots
,X_n)\in{\mathbb{A}}^{\alpha+\sum_{l=1}^n|X_l|}$, \item[$iii)$]
$\omega(X_1,\dots ,X_nz)=\omega(X_1,\dots ,X_n)z$, $\forall\,
z\in{\cal{Z}} ({\mathbb{A}}^\bullet)$, where $(Xz)$ is given in
Proposition \ref{prop7},

\item[$iv)$] $\omega(X_1,X_2,\dots ,X_i,X_{i+1},\dots
,X_n)=(-1)^{|X_i||X_{i+1}|}\omega(X_1,X_2,\dots
,X_{i+1},X_{i},\dots ,X_n)$.
    \end{enumerate}
\end{definition}

\begin{proposition}\label{gradedcalculus}
Let us define
$\Omega_{\text{\textup{Der}}}({\mathbb{A}}^\bullet)=\bigoplus_{n\in{\mathbb{N}},\alpha=0,1}\Omega_{\text{\textup{Der}}}^{n,\alpha}({\mathbb{A}}^\bullet)$.
\par Then, the triplet
$\big(\Omega_{\text{\textup{Der}}}({\mathbb{A}}^\bullet),\times,{\hat{d}}\big)$
is a bigraded differential algebra with the product $\times$ on
$\Omega_{\text{\textup{Der}}}({\mathbb{A}}^\bullet)$ and
differential
${\hat{d}}:\Omega_{\text{\textup{Der}}}^{n,\alpha}({\mathbb{A}}^\bullet)\to\Omega_{\text{\textup{Der}}}^{n+1,\alpha}({\mathbb{A}}^\bullet)$
such that ${\hat{d}}^2=0$ defined by:
\begin{enumerate}\itemsep=0pt
\item[$i)$] For any
$\omega\in\Omega_{\text{\textup{Der}}}^{p,\alpha}({\mathbb{A}}^\bullet)$,
$\eta\in\Omega_{\text{\textup{Der}}}^{q,\alpha}({\mathbb{A}}^\bullet)$
\begin{gather*}
\omega\times\eta(X_1,X_2,\dots ,X_{p+q})\\
\qquad {} ={{1}\over{p!q!}}\sum_{\sigma\in{\mathfrak{S}}(p+q)}
(-1)^{{\textup{sign}}(\sigma)}(-1)^{\sum_{k<l,\sigma(k)>\sigma(l)}|X_k||X_l|+|\eta|\sum_{\sigma(k)\le
p}|X_k|}  \nonumber
\\
\qquad \qquad \quad{}\times \omega(X_{\sigma(1)},\dots
,X_{\sigma(p)})\eta(X_{\sigma(p+1)},\dots ,X_{\sigma(p+q)}).
\end{gather*}
\item[$ii)$] The differential is defined, $\forall\,
\omega\in\Omega_{\text{\textup{Der}}}^{n,\alpha}({\mathbb{A}}^\bullet)$
by:
\begin{gather*}
{\hat{d}}\omega(X_1,X_2,\dots ,X_{n+1})\\
\qquad{}=\sum_{k=1}^{n+1}(-1)^{k+1}(-1)^{|\omega||X_k|}(-1)^{\sum_{n=1}^{k-1}|X_k||X_n|}X_k
\omega(X_1,X_2,\dots ,\vee_k,\dots ,X_{n+1}) \nonumber
\\
\qquad\quad{} +\sum_{1\le k<l\le
n+1}(-1)^{k+l}(-1)^{|X_k||X_l|}(-1)^{\sum_{n=1}^{k-1}|X_k||X_n|}\nonumber
\\
\qquad\qquad\quad\times
(-1)^{\sum_{n=1}^{l-1}|X_l||X_n|}\omega([X_k,X_l]_\bullet,\dots
,\vee_k,\dots \vee_l,\dots ,X_{n+1}),
\end{gather*}
where the symbol $\vee_i$ indicates that $X_i$ is omitted.
\item[$iii)$] A restricted differential calculus is obtained as
follows. Let
${\cal{G}}^\bullet\subset{\text{\textup{Der}}}({\mathbb{A}}^\bullet)$
denote a ${\mathbb{Z}}_2$-graded Lie subalgebra and
${\cal{Z}}({\mathbb{A}}^\bullet)$-submodule. A restricted
differential calculus based on ${\cal{G}}^\bullet$ can be
constructed by replacing in $i)$ and $ii)$
${\text{\textup{Der}}}({\mathbb{A}}^\bullet)$ by
${\cal{G}}^\bullet$ and each
$\Omega_{\text{\textup{Der}}}^{n,\alpha}({\mathbb{A}}^\bullet)$,
$\alpha\in\{0,1\}$, $n\in{\mathbb{N}}$ by the vector space of
$n$-linear maps $\omega$ with homogenous degree $\alpha$,
$\omega:({\cal{G}}^\bullet)^{\times n}\to{\mathbb{A}}^\bullet$.
\end{enumerate}
\end{proposition}
\begin{proof}
The proposition can be verif\/ied from direct calculation.
\end{proof}
As in Section~\ref{section2.2}, we will not write explicitly the
symbol $\times$ for the product of forms as the product to be used
in each case is obvious from the context.

We now def\/ine the connection on ${\mathbb{A}}^\bullet$ and its
curvature as follows:
\begin{definition}\label{connection-curv}
Let
$\Omega_{\text{\textup{Der}}}^{n,\bullet}\equiv\bigoplus_{\alpha=0,1}
\Omega_{\text{\textup{Der}}}^{n,\alpha}({\mathbb{A}}^\bullet)$,
$\forall\, n\in\mathbb{N}$.
\begin{enumerate}\itemsep=0pt
\item[$i)$] The NC connection on ${\mathbb{A}}^\bullet$ is
def\/ined as a linear map of homogeneous degree $0$
$\nabla:\Omega_{\text{\textup{Der}}}^{0,\bullet}\to\Omega_{\text{\textup{Der}}}^{1,\bullet}
$ such that
\begin{equation*}
\nabla(a)={\hat{d}}a+Aa,\qquad \forall\,
a\in{\mathbb{A}}^\bullet,\qquad
A\in\Omega_{\text{\textup{Der}}}^{1,0}({\mathbb{A}}^\bullet),
\end{equation*}
where ${\hat{d}}$ is the dif\/ferential def\/ined in ii) of
Proposition~\ref{gradedcalculus}.
$A\in\Omega_{\text{\textup{Der}}}^{1,0}({\mathbb{A}}^\bullet)$ is
the $1$-form connection.  \par \item[$ii)$] Let $h$ be a Hermitian
structure on ${\mathbb{A}}^\bullet$. The NC connection on
${\mathbb{A}}^\bullet$ is Hermitian if it satisf\/ies
$X(h(a_1,a_2))=h(\nabla_X(a_1),a_2)+(-1)^{|X||a_1|}h(a_1,\nabla_X(a_2))$,
$\forall\, a_1\in{\mathbb{A}}^\bullet$ of homogenous
degree~$|a_1|$, $\forall\, a_2\in{\mathbb{A}}^\bullet$, and for
any real derivation $X$ in
${\text{\textup{Der}}}({\mathbb{A}}^\bullet)$ with homogeneous
deg\-ree~$|X|$. \item[$iii)$] The curvature is def\/ined as a
linear map
$F:\Omega_{\text{\textup{Der}}}^{0,\alpha}({\mathbb{A}}^\bullet)
    \to\Omega_{\text{\textup{Der}}}^{2,\alpha}({\mathbb{A}}^\bullet)$, such that
\begin{equation*}
 F(a)=\nabla(\nabla(a))=({\hat{d}}a+AA)(a),\qquad \forall\, a\in{\mathbb{A}}^\bullet.
\end{equation*}
\end{enumerate}
\end{definition}

Note that the connection def\/ined here preserves the degree of
the ${\mathbb{Z}}_2$-graduation,
\begin{equation*}
\nabla:\Omega_{\text{\textup{Der}}}^{0,\alpha}({\mathbb{A}}^\bullet)
\to\Omega_{\text{\textup{Der}}}^{1,\alpha}({\mathbb{A}}^\bullet).
\nonumber
\end{equation*}
Then, one has for any
$X\in{\text{\textup{Der}}}^\bullet({\mathbb{A}})$ of homogeneous
degree $|X|$, $|A(X)|=|\nabla_X|=|X|$. Notice that we do not use a
NC analogue of a graded connection. This latter appears to have
interesting relationship with the concept of superconnection
introduced by Quillen~\cite{quillen85}. This will be reported
elsewhere \cite{axtjc2}.

From Def\/inition \ref{connection-curv}, one obtains the following
relations that will be used in the sequel
\begin{proposition}\label{lesnabla}
For any $X,Y\in{\text{\textup{Der}}}({\mathbb{A}}^\bullet)$, any
$z\in{\cal{Z}}({\mathbb{A}}^\bullet)$ and any
$a\in{\mathbb{A}}^\bullet $ of homogeneous degree $|X|$, $|Y|$,
$|z|$ and $|a|$, one has
\begin{gather}
\nabla_X(a)=X(a)+A(X)a;\qquad
\nabla_X(ab)=\nabla_X(a)b+(-1)^{|X||a|}aX(b),\qquad \forall \,
b\in{\mathbb{A}}^\bullet,\label{i}
\\
\nabla_{zX}(a)=z\nabla_X(a),\qquad
\nabla_{Xz}(a)=(-1)^{|a||z|}(\nabla_X(a))z, \label{ii}
\\
F(X,Y)(a)=([\nabla_X,\nabla_Y]_\bullet-\nabla_{[X,Y]_\bullet})(a),\qquad
\forall\, a\in{\mathbb{A}}^\bullet . \label{iii}
\end{gather}
Let $^\dag$ denote the involution on ${\mathbb{A}}^\bullet$,
$(ab)^\dag=(-1)^{|a||b|}b^\dag a^\dag$ and let
$h_0(a_1,a_2)=a_1^\dag a_2$ be the Hermitian structure on
${\mathbb{A}}^\bullet$. A NC connection on ${\mathbb{A}}^\bullet$
as given by~Definition~{\rm \ref{connection-curv}} is Hermitian
when for any real derivation $X$ of homogeneous degree, the
$1$-form connection
$A\in\Omega_{\text{\textup{Der}}}^{1,0}({\mathbb{A}}^\bullet)$
satisfies $A(X)^\dag=-A(X)$.
\end{proposition}

\begin{proof}
The relations \eqref{i}--\eqref{iii} can be simply obtained from
standard calculation. Then, assume that the Hermitian structure is
given by $h_0(a_1,a_2)=a_1^\dag a_2$ and consider a Hermitian NC
connection $\nabla$. For any real derivation in
${\text{\textup{Der}}}({\mathbb{A}}^\bullet)$ with homogeneous
degree $|X|$ one has, on one hand,
$X(h_0(a_1,a_2))=X(a_1^\dag)a_2+(-1)^{|X||a_1|}a_1X(a_2)$ and on
the other hand
$h_0(\nabla_X(a_1),a_2)+(-1)^{|X||a_1|}h_0(a_1,\nabla_X(a_2))=(X(a_1^\dag)+(-1)^{|X||a_1|}a_1^\dag
A^\dag(X))a_2+(-1)^{|X||a_1|}a_1^\dag(X(a_2)+A(X)a_2)$, using
Proposition~\ref{lesnabla} and the reality of $X$. Then, the
relation given in Def\/inition~\ref{connection-curv} \[
X(h_0(a_1,a_2)){\cal{A}}(X)=h_0(\nabla_X(a_1),a_2)+(-1)^{|X||a_1|}h_0(a_1,\nabla_X(a_2))
 \]
 is fulf\/illed provided $A^\dag(X)=-A(X)$.
\end{proof}

The gauge transformations are def\/ined in a way which preserves
the degree of the NC connection. In the following, we will focus
on unitary gauge transformations that are compatible with the
Hermitian structure.

\begin{definition}
The gauge group is def\/ined as
${\textup{Aut}}^0({\mathbb{A}}^\bullet)$, the group of the
automorphisms with degree $0$ of ${\mathbb{A}}^\bullet$. Let
$a^\varphi\equiv\varphi(a)$ $\forall\,
\varphi\in{\textup{Aut}}^0({\mathbb{A}}^\bullet)$, $\forall
a\in{\mathbb{A}}^\bullet$. Let $h$ be a Hermitian structure on
${\mathbb{A}}^\bullet$. The unitary gauge group is def\/ined as
the subgroup
${\cal{U}}({\mathbb{A}}^\bullet)\subset{\textup{Aut}}^0({\mathbb{A}}^\bullet)$
such that
${\cal{U}}({\mathbb{A}}^\bullet)=\{\varphi\in{\textup{Aut}}^0({\mathbb{A}}^\bullet)\,
/\, h(a_1^\varphi,a_2^\varphi)=h(a_1,a_2),\ \forall\,
a_1,a_2\in{\mathbb{A}}^\bullet\}$.
\end{definition}

When $h(a_1,a_2)=h_0(a_1,a_2)=a_1^\dag a_2$, it follows from this
def\/inition that the gauge group
${\cal{U}}({\mathbb{A}}^\bullet)$ can be identif\/ied obviously
with the group of unitary elements of ${\mathbb{A}}^\bullet$.

\begin{proposition} \label{graded-gauge}
For any NC connection $\nabla$ as given in Definition~{\rm
\ref{connection-curv}} and for any
$\varphi\in{\textup{Aut}}^0({\mathbb{A}}^\bullet)$, the linear map
of homogeneous degree $0$
$\nabla^\varphi:\Omega_{\text{\textup{Der}}}^{0,\bullet}\to\Omega_{\text{\textup{Der}}}^{1,\bullet}
$ such that
$\nabla^\varphi(a)(X)=(\varphi^{-1}\circ\nabla\circ\varphi)(a)(X)$,
$\forall\, a\in{\mathbb{A}}^\bullet$, $\forall \,
X\in{\text{\textup{Der}}}({\mathbb{A}}^\bullet)$ defines a NC
connection. The corresponding $1$-form connection is defined by
$A^\varphi=\varphi^{-1}({\mathbb{I}}){\hat{d}}\varphi({\mathbb{I}})+\varphi^{-1}({\mathbb{I}})A\varphi({\mathbb{I}})$.
The gauge transformation of the curvature is given by
$F^{\varphi}(a)=(\varphi^{-1}\circ F\circ\varphi)(a)$, $\forall \,
a\in{\mathbb{A}}^\bullet$.
\end{proposition}

\begin{proof}
First, as for $iii)$ of Proposition~\ref{prop5}, one has
$a^\varphi=\varphi(a)=\varphi({\mathbb{I}})a$,
$\forall\varphi\in{\textup{Aut}}^0({\mathbb{A}}^\bullet)$. To
simplify the notations, we set $g=\varphi({\mathbb{I}})$. Then,
from Def\/inition~\ref{connection-curv}, one obtains
$\varphi^{-1}(\nabla(\varphi(a)))=g^{-1}\nabla(ga)=g^{-1}{\hat{d}}(ga)+g^{-1}Aga
={\hat{d}}a+(g^{-1}{\hat{d}}g+g^{-1}Ag)a$. The expression for
$A^{\varphi}$ follows. The gauge transformation for the curvature
can be obtained by a simple calculation.
\end{proof}

Proposition~\ref{prop2.9} and Lemma~\ref{lemma} must be slightly
modif\/ied to take into account the total grading
${\mathbb{N}}\times{\mathbb{Z}}_2$. In fact, this can be compactly
expressed in terms of dif\/ferential forms. Note that the bracket
on $\Omega_{\text{\textup{Der}}}({\mathbb{A}}^\bullet)$ is
$[\omega_p,\eta_q]_\bullet=\omega\eta-(-1)^{pq}(-1)^{|\omega||\eta|}\eta\omega$,
$\forall\,
\omega\in\Omega_{\text{\textup{Der}}}^{p,|\omega|}({\mathbb{A}}^\bullet)$,
$\forall \,
\eta\in\Omega_{\text{\textup{Der}}}^{q,|\eta|}({\mathbb{A}}^\bullet)$.
The following lemma holds

\begin{lemma}
Assume that
$\exists\,\eta\in\Omega_{\text{\textup{Der}}}^{1,0}({\mathbb{A}}^\bullet)\,
/\, {\hat{d}}a=[\eta,a]_\bullet=\eta a-a\eta$, $\forall\,
a\in{\mathbb{A}}^\bullet$ where ${\hat{d}}$ is the differential
given in Proposition~{\rm \ref{gradedcalculus}}. Then, one has the
properties:
\begin{enumerate}\itemsep=0pt
\item[$i)$] The map $\nabla^{\rm
inv}:\Omega_{\text{\textup{Der}}}^{0,\alpha}({\mathbb{A}}^\bullet)
    \to\Omega_{\text{\textup{Der}}}^{1,\alpha}({\mathbb{A}}^\bullet)$, $\alpha=0,1$ such that $\nabla^{\rm inv}(a)={\hat{d}}a-\eta a=-a\eta$, $\forall\,  a\in{\mathbb{A}}^\bullet$, defines a connection in the sense of Definition~{\rm \ref{connection-curv}}. It is called the canonical connection. $\nabla^{\rm inv}$ is gauge invariant: $(\nabla^{\rm inv})^g(a)=\nabla^{\rm inv}(a)$, $\forall\,  g\in{\cal{U}}({\mathbb{A}}^\bullet)$.
\item[$ii)$] The curvature for the canonical connection is defined
by the linear map $F^{\rm
inv}:\Omega_{\text{\textup{Der}}}^{0,\alpha}({\mathbb{A}}^\bullet)\to\Omega_{\text{\textup{Der}}}^{2,\alpha}
    ({\mathbb{A}}^\bullet)$ such that $F^{\rm inv}(a)=(-{\hat{d}}\eta+\eta\eta)(a)$, $\forall \, a\in{\mathbb{A}}^\bullet$.

\item[$iii)$] For any NC connection $\nabla$ in the sense of
Definition~{\rm \ref{connection-curv}},
${\cal{A}}\in\Omega_{\text{\textup{Der}}}^{1,0}({\mathbb{A}}^\bullet)$
such that ${\cal{A}}(a)=(\nabla-\nabla^{\rm inv})(a)=(A+\eta)(a)$,
$\forall\, a\in{\mathbb{A}}^\bullet$ defines a tensor form.
${\cal{A}}(X)$, $\forall\,
X\in{\text{\textup{Der}}}({\mathbb{A}}^\bullet)$ are called the
covariant coordinates. The curvature for any NC connection
$\nabla$ is defined by the linear map
$F:\Omega_{\text{\textup{Der}}}^{0,\alpha}({\mathbb{A}}^\bullet)
    \to\Omega_{\text{\textup{Der}}}^{2,\alpha}({\mathbb{A}}^\bullet)$ such that $F(a)=F^{\rm inv}(a)+({\cal{A}}{\cal{A}}+{\hat{d}}{\cal{A}}-[{\cal{A}},\eta]_\bullet))(a)$, $\forall\, a\in{\mathbb{A}}^\bullet$.
\item[$iv)$] For any
$X,Y\in{\text{\textup{Der}}}({\mathbb{A}}^\bullet)$, $\forall\,
a\in{\mathbb{A}}^\bullet$ with respective homogeneous degree
$|X|$, $|Y|$, $|a|$, the following relations hold:
\begin{gather}
\nabla^{\rm inv}_X(a)=-(-1)^{|X||a|}a\eta(X),\qquad
\nabla_X(a)={\cal{A}}(X)a-(-1)^{|X||a|}a\eta(X) ,
\label{eq:formule1}
\\
F(X,Y)=[{\cal{A}}(X),{\cal{A}}(Y)]_\bullet-{\cal{A}}([X,Y]_\bullet)+(\eta([X,Y]_\bullet)
-[\eta(X),\eta(Y)]_\bullet),\label{eq:formule2}
\\
F^{\rm
inv}(X,Y)=-[\eta(X),\eta(Y)]_\bullet+\eta([X,Y]_\bullet).\label{eq:formule3}
\end{gather}
\end{enumerate}
\end{lemma}

\begin{proof}
From Def\/inition \ref{connection-curv}, any $1$-form in
$\Omega_{\text{\textup{Der}}}^{1,0}({\mathbb{A}}^\bullet)$
def\/ines a connection. In particular, pick $A=-\eta$. The
property $i)$ follows. Then, $\forall \,
g\in{\cal{U}}({\mathbb{A}}^\bullet)$, $(\nabla^{\rm
inv})^g(a)=g^\dag\nabla^{\rm
inv}(ga)=-g^\dag(ga\eta)=-a\eta=\nabla^{\rm inv}(a)$. The
corresponding curvature is determined by the $2$-form
$F={\hat{d}}A+AA$ for $A=-\eta$. So, $F^{\rm
inv}=-{\hat{d}}\eta+\eta\eta$. The property for ${\cal{A}}$ in
$iii)$ is a direct consequence of the def\/inition of a tensor
form. The expression for $F$ in terms of ${\cal{A}}$ is obtained
by setting $A={\cal{A}}-\eta$ in
Def\/inition~\ref{connection-curv} and using the expression for
$F^{\rm inv}$. Finally, the relations given in $iv)$ are obtained
by using the main properties of the graded dif\/ferential calculus
given in Proposition~\ref{gradedcalculus}. One has in particular
${\cal{A}}{\cal{A}}(X,Y)=[{\cal{A}}(X),{\cal{A}}(Y)]_\bullet$,
$[{\cal{A}},\eta]_\bullet(X,Y)=[\eta(X),{\cal{A}}(Y)]_\bullet-(-1)^{|X||Y|}[\eta(Y),{\cal{A}}(X)]_\bullet$,
${\hat{d}}{\cal{A}}(X,Y)=[\eta(X),{\cal{A}}(Y)]_\bullet-(-1)^{|X||Y|}[\eta(Y),{\cal{A}}(X)]_\bullet-{\cal{A}}([X,Y]_\bullet)$,
${\hat{d}}\eta(X,Y)=2[\eta(X),\eta(Y)]_\bullet-\eta([X,Y]_\bullet)$.
Then, the relations \eqref{eq:formule1}--\eqref{eq:formule3}
follow.
\end{proof}

We will now apply the above non-graded and graded dif\/ferential
calculi based on the derivations of associative algebras to two
cases. The f\/irst case deals with a non-graded dif\/ferential
calculus on the Moyal algebra ${\cal{M}}$ stemming from the
maximal subalgebra of ${\text{\textup{Der}}}({\cal{M}})$ whose
elements can be interpreted as inf\/initesimal symplectomorphisms.
This gives rise to a natural construction of Yang--Mills--Higgs
type models def\/ined on ${\cal{M}}$. The second case deals with
a~${\mathbb{Z}}_2$-graded dif\/ferential calculus on a
${\mathbb{Z}}_2$-graded associative algebra built from two copies
of ${\cal{M}}$. Due to the grading, the gauge theory built from
the square of the corresponding curvatures involves as
contributions the action~\eqref{eq:decadix1} derived in~\cite{de
Goursac:2007gq} and \cite{Grosse:2007qx} as well the
renormalisable NC $\varphi^4$-model with harmonic term elaborated
in \cite{Grosse:2004yu,Grosse:2003aj}.

From now on, the Hermitian structure is $h_0(a_1,a_2)=a_1^\dag
a_2$. According to the above discussion, the gauge group is the
group of unitary elements of the associative algebra.\par

\section{Application: gauge theories on the Moyal algebra}\label{section3}

\subsection{General properties of the Moyal algebra}\label{section3.1}

In this subsection, we collect the properties of the Moyal algebra
that will be used in the sequel. For more details, see
e.g.~\cite{Gracia-Bondia:1987kw,Varilly:1988jk}). Let
${\cal{S}}({\mathbb{R}}^D)\equiv{\cal{S}}$ and
${\cal{S}}^\prime({\mathbb{R}}^D)\equiv{\cal{S}}^\prime$, with
$D=2n$,  be respectively the space of complex-valued Schwartz
functions on ${\mathbb{R}}^D$ and the dual space of tempered
distributions on ${\mathbb{R}}^D$. The complex conjugation in
${\cal{S}}$, $a\mapsto a^\dag$, $\forall\, a\in{\cal{S}}$,
def\/ines a natural involution in ${\cal{S}}$ that can be extended
to ${\cal{S}}^\prime$ by duality and that will be used in the rest
of this paper. Let $\Theta_{\mu\nu}$ be an invertible constant
skew-symmetric matrix which can be written as
$\Theta=\theta{{\Sigma}}$ where $\Sigma$ is the ``block-diagonal''
matrix, ${{\Sigma}}={\rm diag}(J,\dots ,J)$ involving $n$
$(2\times2)$ mat\-rix~$J$ given by $J =\begin{pmatrix} 0&-1 \\ 1&
0 \end{pmatrix}$ and the parameter $\theta$ has mass dimension
$-2$. We use the notation $y\Theta^{-1}z\equiv y_\mu
\Theta^{-1}_{\mu\nu}z^\nu$. The following proposition summarises
properties relevant for the ensuing discussion:

\begin{proposition}
Let the $\star$-Moyal product be defined as
$\star:{\cal{S}}\times{\cal{S}}\to{\cal{S}}$ by
\begin{equation}
(a\star b)(x)=\frac{1}{(\pi\theta)^D}\int d^Dyd^Dz\
a(x+y)b(x+z)e^{-i2y\Theta^{-1}z},\qquad \forall \,
a,b\in{\cal{S}}.\label{eq:moyal}
\end{equation}
Then, $({\cal{S}},\star)$ is a non unital associative involutive
Fr\'echet algebra with faithful trace given by $\int d^Dx\ (a\star
b)(x)=\int d^Dx$ $(b\star a)(x)=\int d^Dx  a(x)\cdot b(x)$, where
the symbol ``$\cdot$'' is the usual commutative product of functions in
${\cal{S}}$.
\end{proposition}

The $\star$ product~\eqref{eq:moyal} can be further extended to
${\cal{S}}^\prime\times{\cal{S}}\to{\cal{S}}^\prime$ upon using
duality of vector spaces: $\langle T\star a,b \rangle = \langle
T,a\star b\rangle$, $\forall\, T\in{\cal{S}}^\prime$, $\forall\,
a,b\in{\cal{S}}$. In a similar way, \eqref{eq:moyal} can be
extended to ${\cal{S}} \times
{\cal{S}}^\prime\to{\cal{S}}^\prime$, via $\langle a\star T,b
\rangle = \langle T,b\star a\rangle$, $\forall\,
T\in{\cal{S}}^\prime$, $\forall\, a,b\in{\cal{S}}$. Then, the
Moyal algebra is def\/ined as \cite{Gracia-Bondia:1987kw,
Varilly:1988jk}.

\begin{definition}\label{moyaldef}
Let ${\cal{M}}_L$ and ${\cal{M}}_R$ be respectively def\/ined by
${\cal{M}}_L=\{T\in{\cal{S}}^\prime\, /\, a\star T\in{\cal{S}},\
\forall\, a\in{\cal{S}}\}$ and
${\cal{M}}_R=\{T\in{\cal{S}}^\prime\, /\, T\star a\in{\cal{S}},\
\forall \, a\in{\cal{S}}\}$. Then, the Moyal algebra ${\cal{M}}$
is def\/ined by
\begin{gather*}
{\cal{M}}={\cal{M}}_L\cap{\cal{M}}_R.
\end{gather*}
\end{definition}
Notice that ${\cal{M}}_L$ and ${\cal{M}}_R$ are sometimes called
in the literature respectively the left and right multiplier
algebras. By construction, ${\cal{S}}$ is a two-sided ideal of
${\cal{M}}$. The essential structural properties of the Moyal
algebra that we will need are summarised in the following
proposition:

\begin{proposition}
${\cal{M}},\star$ is a maximal unitisation of $({\cal{S}},
\star)$. It is a locally convex associative unital $*$-algebra. It
involves the plane waves, the Dirac distribution and its
derivatives and the polynomial functions.
\end{proposition}

In the following, we will need the following asymptotic formula
for the $\star$-product
\begin{gather}
(a\star b)(x)=a(x)\cdot b(x)\nonumber\\
\phantom{(a\star b)(x)=}{}
+\sum_{n=1}^\infty{{1}\over{n!}}\left({{i}\over{2}}\Theta^{\mu_1\nu_1}{{\partial}\over{\partial
x^{\mu_1}}} {{\partial}\over{\partial y^{\nu_1}}}\right)\dots
\left({{i}\over{2}}\Theta^{\mu_n\nu_n}{{\partial}\over{\partial
x^{\mu_n}}}{{\partial}\over{\partial
y^{\nu_n}}}\right)a(x)b(y)\vert_{x=y}, \label{eq:moyal-as}
\end{gather}
which holds for any polynomial functions $a$, $b$. For a detailed
study of the validity of~\eqref{eq:moyal-as}, see
e.g.~\cite{asymp-valid}.

\begin{proposition}\label{center-inner}
The center is ${\cal{Z}}({\cal{M}})={\mathbb{C}}$.
\end{proposition}

Other relevant properties of the $\star$-product that hold on
${\cal{M}}$ and will be used in the sequel are
\begin{proposition}
For any $a,b\in{\cal{M}}$, one has the following relations on
${\cal{M}}$:
\begin{subequations}\label{3.4}
\begin{gather}
\partial_\mu(a\star b)=\partial_\mu a\star b+a\star\partial_\mu
b,\qquad (a\star b)^\dag=b^\dag\star a^\dag,\qquad
[x_\mu,a]_\star=i\Theta_{\mu\nu}\partial_\nu a, \label{eq:relat1}
\\
x_\mu\star a=(x_\mu\cdot
a)+{{i}\over{2}}\Theta_{\mu\nu}\partial_\nu a, \qquad x_\mu(a\star
b)=(x_\mu.a)\star b
-{{i}\over{2}}\Theta_{\mu\nu}a\star\partial_\nu b
,\label{eq:relat2}
\\
(x_\mu\cdot x_\nu)\star a=x_\mu\cdot x_\nu\cdot
a+{{i}\over{2}}(x_\mu\Theta_{\nu\beta}+x_\nu\Theta_{\mu\beta})\partial_\beta
a
-{{1}\over{4}}\Theta_{\mu\alpha}\Theta_{\nu\sigma}\partial_\alpha\partial_\sigma
a, \label{eq:relat3}
\\
a\star(x_\mu\cdot x_\nu)=x_\mu\cdot x_\nu\cdot
a-{{i}\over{2}}(x_\mu\Theta_{\nu\beta}+x_\nu\Theta_{\mu\beta})\partial_\beta
a
-{{1}\over{4}}\Theta_{\mu\alpha}\Theta_{\nu\sigma}\partial_\alpha\partial_\sigma
a, \label{eq:relat4}
\\
[(x_\mu\cdot x_\nu\cdot x_\rho),a]_\star=i(x_\rho
x_\mu\Theta_{\nu\beta}+x_\nu x_\rho\Theta_{\mu\beta}+x_\mu
x_\nu\Theta_{\rho\beta})\partial_\beta
a-\!{{i}\over{4}}\Theta_{\mu\alpha}\Theta_{\nu\sigma}\Theta_{\rho\lambda}\partial_\alpha\partial_\sigma\partial_\lambda
a).\! \!\!\!\!\label{eq:relat5}
\end{gather}
\end{subequations}
\end{proposition}

\begin{proof}
The relations can be obtained by calculations, using from instance
the asymptotic expansion \eqref{eq:moyal-as}.
\end{proof}
Notice that, as a special case of  the last relation
\eqref{eq:relat1}, one obtains the celebrated relation among the
``coordinate functions'' def\/ined on~${\cal{M}}$:
\begin{equation}
[x_\mu,x_\nu]_\star =i\Theta_{\mu\nu}, \label{eq:comrelation}
\end{equation}
where we set $[a,b]_\star\equiv a\star b-b\star a$.

As a f\/inal remark, note that the Moyal algebra has
${\cal{Z}}({\cal{M}})={\mathbb{C}}$ as trivial center, which stems
from the fact that  $\Theta_{\mu\nu}$ is non degenerate. This
simplif\/ies the situation regarding all the structures of modules
over ${\cal{Z}}({\cal{M}})$ that are involved in the present
algebraic scheme. In the present case, these are simply replaced
by vector spaces over~${\mathbb{C}}$.

\subsection[Differential calculus and inner derivations]{Dif\/ferential calculus and inner derivations}\label{map-eta}

The vector space of derivations of ${\cal{M}}$ is inf\/inite
dimensional. Then, a dif\/ferential calculus based on the full
derivation algebra ${\text{\textup{Der}}}({\cal{M}})$ would give
rise to gauge potentials with an inf\/inite number of components.
In view of the construction of physically oriented gauge theories
on Moyal spaces, it is more convenient to deal with gauge
potentials having a f\/inite number of components. These occur
within restricted dif\/ferential calculi based on Lie subalgebras
of ${\text{\textup{Der}}}({\mathbb{A}})$, as given in
Proposition~\ref{prop3}. In the following, we will consider two
Lie subalgebras of ${\text{\textup{Der}}}({\cal{M}})$, denoted
by~${\cal{G}}_1$ and ${\cal{G}}_2$. The f\/irst one is Abelian and
is simply related to the ``spatial derivatives'' $\partial_\mu$.
The resulting dif\/ferential calculus underlies almost all the
constructions of NCFT def\/ined on Moyal spaces. For further
convenience, we set from now on
\begin{equation}
\partial_\mu a=[i\xi_\mu,a]_\star, \qquad \xi_\mu=-\Theta^{-1}_{\mu\nu}x^\nu,\qquad\forall\, a\in {\cal{M}} . \label{eq:innerxi}
\end{equation}
The second derivation Lie subalgebra ${\cal{G}}_2$, such that
${\cal{G}}_1\subset{\cal{G}}_2$, is the maximal subalgebra of
${\text{\textup{Der}}}({\cal{M}})$ whose derivations can be
interpreted as inf\/initesimal symplectomorphisms. Notice that in
each case, the existence of NC connections is explicitly
verif\/ied, due to the existence of canonical gauge invariant
connections of the type given in Proposition~\ref{prop2.9}.

\begin{proposition}\label{propP2}
Let ${\cal{P}}_2\subset{\cal{M}}$ denote the set of polynomial
functions with degree $d\le2$. Let
$\{a,b\}_{PB}\equiv\Theta_{\mu\nu}{{\partial a}\over{\partial
x_\mu}}{{\partial b}\over{\partial x_\nu}}$ for any polynomial
function $a,b\in{\cal{M}}$ denote the Poisson bracket for the
symplectic structure defined by $\Theta_{\mu\nu}$. Then,
${\cal{P}}_2$ equipped with the Moyal bracket $[~,~]_\star$ is
a~Lie algebra which satisfies
\begin{gather*}
[P_1,P_2]_\star=i\{P_1,P_2\}_{PB},\qquad \forall\, P_1,P_2\in{\cal{P}}_2. 
\end{gather*}
\end{proposition}

\begin{proof}
Using \eqref{eq:moyal-as}, one infers that $(P_1\star P_2)(x)$,
$\forall\, P_1,P_2\in{\cal{P}}_2$ truncate to a f\/inite
expansion. Namely,  $(P_1\star P_2)(x)=P_1(x)\cdot
P_2(x)+{{i}\over{2}}\Theta_{\mu\nu}{{\partial P_1}\over{\partial
x_\mu}}{{\partial P_2}\over{\partial
x_\nu}}-{{1}\over{4}}\Theta_{\mu\nu}\Theta_{\rho\sigma}{{\partial^2
P_1}\over{\partial x_\mu\partial x_\rho}}{{\partial
P_2}\over{\partial x_\nu\partial x_\sigma}}$ where the last term
is a constant. Then, $[P_1,P_2]_\star=i\Theta_{\mu\nu}{{\partial
P_1}\over{\partial x_\mu}}{{\partial P_2}\over{\partial x_\nu}}$
from which follows the proposition.
\end{proof}

Consider now the Lie subalgebra
${\cal{G}}_2\subset{\text{\textup{Der}}}({\cal{M}})$ which is the
image of ${\cal{P}}_2$ by ${\textup{Ad}}$,
${\cal{G}}_2=\{X\in{\text{\textup{Der}}}({\cal{M}}\, /\,
X={\textup{Ad}}_P,\ P\in{\cal{P}}_2\}$. In order to apply
Proposition~\ref{prop2.9} and Lemma~\ref{lemma} to the present
situation, one has to def\/ine properly the 1-form $\eta$ from
which most of the objects entering the construction of gauge
theories are derived. To do this, one def\/ines the linear map
$\eta$ as
\begin{equation}
\eta:  \ \ {\cal{G}}_2\to{\cal{P}}_2\quad /\quad \eta(X)=P-P(0),
\qquad  \forall\, X\in{\cal{G}}_2 , \label{eq:thetaform}
\end{equation}
where $P(0)\in{\mathbb{C}}$ is the evaluation of the polynomial
function $P$ at $x=0$. Then,
$X(a)={\textup{Ad}}_P(a)={\textup{Ad}}_{\eta(X)}(a)$, $\forall\,
X\in{\cal{G}}_2$, $\forall\, a\in{\cal{M}}$ and
\eqref{eq:thetaform} can be used to def\/ine the 1-form $\eta$ in
the present case. Notice that $\eta$ does not def\/ine a morphism
of Lie algebra since, as implied by the property $i)$ of
Lemma~\ref{lemma}, one has
$\eta([X_1,X_2])-[\eta(X_1),\eta(X_2)]\in{\mathbb{C}}$ and is non
zero in the present case. Indeed, take for instance
$\eta(\partial_\mu)=i\xi_\mu$ (see~\eqref{eq:innerxi}); then
$\eta([\partial_\mu,\partial_\nu])-[\eta(\partial_\mu),\eta(\partial_\nu)]_\star=[\xi_\mu,\xi_\nu]_\star
=-i\Theta^{-1}_{\mu\nu}$.

At this point, some remarks are in order. Proposition~\ref{propP2}
singles out two subalgebras of derivations, whose elements are
related to inf\/initesimal symplectomorphisms. These are sometimes
called area-preserving dif\/feomorphims in the physics literature.
The f\/irst algebra ${\cal{G}}_1$ is Abelian and is simply the
image by ${\textup{Ad}}$ of the algebra generated by
$\{P_\mu=x_\mu,\ \mu=1,\dots ,D\}$, i.e.\ the polynomials with
degree~$1$. It is the algebra related to the spatial derivatives
$\partial_\mu$ in view of the 3rd relation of~\eqref{eq:relat1}
and~\eqref{eq:innerxi}. One has immediately, due
to~\eqref{eq:comrelation}
\begin{equation*}
[\partial_\mu,\partial_\nu](a)=[{\textup{Ad}}_{i\xi_\mu},{\textup{Ad}}_{i\xi_\nu}](a)
={\textup{Ad}}_{[i\xi_\mu,i\xi_\nu]_\star}(a)=0, \qquad \forall\, a\in{\cal{M}}. 
\end{equation*}
Note that the interpretation of $[x_\mu,a]_\star$ as the Lie
derivative along the (constant) Hamiltonian vector f\/ield
$i\Theta_{\mu\nu}$ is obvious. The dif\/ferential calculus based
on ${\cal{G}}_1$ is the minimal one that can be set-up on the
Moyal algebra and actually underlies most of the studies of the
NCFT on Moyal spaces. The second algebra ${\cal{G}}_2$ is the
image by ${\textup{Ad}}$ of $\{P_{\mu\nu}=(x_\mu x_\nu),\
\mu,\nu=1,\dots ,D\}$, the algebra generated by the polynomials
with degree~2. It is the maximal subalgebra of
${\text{\textup{Der}}}({\cal{M}})$ whose elements can be related
to symplectomorphims. Observe that from~\eqref{eq:relat3}
and~\eqref{eq:relat4} one has
\begin{equation*}
[(x_\mu\cdot x_\nu),a]_\star=i(x_\mu\Theta_{\nu\beta}+x_\nu\Theta_{\mu\beta})\partial_\beta a,
\end{equation*}
so that the bracket in the LHS can again be interpreted as the Lie
derivative along a Hamiltonian vector f\/ield. Note that this is
no longer true for polynomials with degree $d \ge 3$, which is
apparent from~\eqref{eq:relat5} for $d = 3$.

Once ${\cal{G}}_1$ or ${\cal{G}}_2$ is choose and the
corresponding 1-form $\eta$ is determined, all the properties and
mathematical status of the various objects entering the
construction of gauge theories on Moyal spaces are entirely
f\/ixed from Proposition~\ref{prop2.9} and Lemma~\ref{lemma}. The
corresponding relations are summarised below for further
convenience. For any $X\in{\cal{G}}_i$, $i=1,2$, one has
\begin{gather}
\nabla^{\rm inv}_X(a)=-a\star\eta(X),\qquad
{\cal{A}}(X)=A(X)+\eta(X), \label{eq:final1}
\\
 \nabla_X(a)=\nabla^{\rm inv}_X(a)+{\cal{A}}_X\star a=\nabla^{\rm inv}_X(a)+(A(X)+\eta(X))\star a=X(a)+A(X)\star a,\label{eq:final2}
\\
{F}{(X,Y)}=([{\cal{A}}(X),{\cal{A}}(Y)]-{\cal{A}}{[X,Y]})-([\eta(X),\eta(Y)]-\eta{[X,Y]}).\label{eq:final3}
\end{gather}

\subsection{Application to the Moyal space}\label{moyalcase}

Consider f\/irst the Abelian algebra ${\cal{G}}_1$ generated by
the spatial derivatives~\eqref{eq:innerxi}. Then, after doing a
simple rescaling $A(X)\to-iA(X)$ (i.e.\ def\/ining
$\nabla_X({\mathbb{I}})\equiv-iA(X)$ so that Hermitian connections
satisfy $A(X)^\dag=A(X)$ for any real derivation $X$) in order to
make contact with the notations of e.g.~\cite{de
Goursac:2007gq,Wallet:2007 em}, and def\/ining
$\eta(\partial_\mu)\equiv\eta_\mu$, ${\cal{A}}(\partial_\mu)\equiv
{\cal{A}}_\mu$, $A(\partial_\mu)\equiv A_\mu$,
$F(\partial_\mu,\partial_\nu)\equiv F_{\mu\nu}$, $\mu=1,\dots ,D$,
a straightforward application of~\eqref{eq:thetaform}, and
\eqref{eq:final1}--\eqref{eq:final3} yields
\begin{gather*}
\eta_\mu=i\xi_\mu,\qquad \nabla^{\rm
inv}_\mu(a)=-ia\star\xi_\mu,\qquad \forall\,
a\in{\cal{M}},\nonumber
\\
{\cal{A}}_\mu=-i(A_\mu-\xi_\mu),\qquad \nabla_\mu(a)=-ia\star\xi_\mu+{\cal{A}}_\mu\star a=\partial_\mu a-iA_\mu\star a,\qquad  \forall\,  a\in{\cal{M}}, 
\\
F_{\mu\nu}=[{\cal{A}}_\mu,{\cal{A}}_\nu]_\star-i\Theta^{-1}_{\mu\nu}=-i\big(\partial_\mu A_\nu-\partial_\nu A_\mu-i[A_\mu,A_\nu]_\star\big), 
\end{gather*}
which f\/ix the respective mathematical status of the objects
involved in most of the studies of NCFT on Moyal spaces. The group
of (unitary) gauge transformation is the group of unitary elements
of ${\cal{M}}$, ${\cal{U}}({\cal{M}})$, as def\/ined in
Section~\ref{section2} and one has
\begin{gather*}
A_\mu^g=g\star A_\mu\star g^\dag+ig\star \partial_\mu
g^\dag,\qquad  {\cal{A}}_\mu^g=g\star {\cal{A}}_\mu\star g^\dag,
\\  F_{\mu\nu}^g=g\star F_{\mu\nu}\star g^\dag,\qquad \forall\, g\in{\cal{U}}({\cal{M}}).
\end{gather*}
Consider now the algebra ${\cal{G}}_2$. Let
${\bar{{\cal{G}}}}_2\subset {\cal{G}}_2$ denote the subalgebra of
${\cal{G}}_2$ whose image in ${\cal{M}}$ by the map~$\eta$
\eqref{eq:thetaform} corresponds to the monomials of degree $2$.
The image involves ${{D(D+1)}\over{2}}$ elements def\/ined by
\begin{equation*}
\eta(X_{\mu\nu})=i\xi_\mu\xi_\nu\equiv\eta_{(\mu\nu)},\qquad
\forall\,  X_{\mu\nu}\in{\bar{{\cal{G}}}}_2,\qquad \forall\,
\mu,\nu=1,\dots ,D.
\end{equation*}
The symbol $(\mu\nu )$ denotes symmetry under the exchange
$\mu\leftrightarrow\nu$. Notice that the def\/inition
for~$\eta_{\mu\nu}$ corresponds to real derivations. One has
\begin{equation}
[\eta_{(\mu\nu)},\eta_{(\rho\sigma)}]_\star=-\big(\Theta^{-1}_{\rho\nu}{\eta_{(\mu\sigma)}}
+\Theta^{-1}_{\sigma\nu}{\eta_{(\mu\rho)}}
+\Theta^{-1}_{\rho\mu}{\eta_{(\nu\sigma)}}+\Theta^{-1}_{\sigma\mu}{\eta_{(\nu\rho)}}\big).\label{eq:slnr}
\end{equation}
which def\/ine the generic commutation relations for the
$sp(2n,{\mathbb{R}})$ algebra. Then, the algebra~${\cal{G}}_2$ we
choose is generated by~$\{\partial_\mu, X_{\mu\nu}\}$. Its image
in ${\cal{M}}$ by the map $\eta$ \eqref{eq:thetaform} is the
algebra $isp(2n,{\mathbb{R}})$. One has the additional commutation
relations
\begin{equation}
[\eta_\mu,\eta_{(\rho\sigma)}]_\star=\big(\Theta^{-1}_{\mu\rho}\eta_\sigma+\Theta^{-1}_{\mu\sigma}\eta_\rho\big).
\label{eq:addicom}
\end{equation}
Notice that any derivation related to $isp(2n,{\mathbb{R}})$ can
be viewed as generating an inf\/initesimal symplectomorphism, as
discussed above. Accordingly, the subalgebra
${\cal{G}}_1\subset{\cal{G}}_2$ can actually be interpreted
physically as corresponding to spatial directions while
${\bar{{\cal{G}}}}_2$ corresponds to (symplectic) rotations.
Notice also that in the case $D=2$, upon def\/ining
\begin{equation*}
\eta_{X_1}={{i}\over{4{\sqrt{2}}\theta}}\big(x_1^2+x_2^2\big),\qquad
\eta_{X2}={{i}\over{4{\sqrt{2}}\theta}}\big(x_1^2-x_2^2\big),\qquad
\eta_{X3}={{i}\over{2{\sqrt{2}}\theta}}(x_1x_2) 
\end{equation*}
one would obtain the following commutation relations
\begin{gather*}
[\eta_{X1},\eta_{X2}]_\star={{1}\over{{\sqrt{2}}}}\eta_{X3},\qquad
[\eta_{X2},\eta_{X3}]_\star=-{{1}\over{{\sqrt{2}}}}\eta_{X1},\qquad
[\eta_{X3},\eta_{X1}]_\star={{1}\over{{\sqrt{2}}}}\eta_{X2}, \nonumber
\\
[\eta_1,\eta_{X1}]_\star={{1}\over{2{\sqrt{2}}}}\eta_2 ,\qquad
[\eta_2,\eta_{X1}]_\star=-{{1}\over{2{\sqrt{2}}}}\eta_1,
\\
[\eta_1,\eta_{X2}]_\star={{1}\over{2{\sqrt{2}}}}\eta_2 ,\qquad
[\eta_2,\eta_{X2}]_\star={{1}\over{2{\sqrt{2}}}}\eta_1,
\\
[\eta_1,\eta_{X3}]_\star=-{{1}\over{2{\sqrt{2}}}}\eta_1 ,\qquad
[\eta_2,\eta_{X3}]_\star={{1}\over{2{\sqrt{2}}}}\eta_2,
\end{gather*}
therefore making contact with the work carried out
in~\cite{italiens}. Note that \cite{italiens} did not consider the
construction of gauge theories on Moyal spaces but was only
focused on the construction of subalgebras of the $D=4$ Moyal
algebra starting from a set of constraints forming a~(subalgebra
of~a)~$sp(2n,{\mathbb{R}})$ algebra and the obtention of the
algebra of smooth functions of ${\mathbb{R}}^3$ from a~commutative
limit.

A direct application of \eqref{eq:thetaform}, and
\eqref{eq:final1}--\eqref{eq:final3} permits one to determine the
invariant connection and the tensor form. One obtains
\begin{gather}
\nabla^{\rm inv}(\partial_\mu)(a)\equiv\nabla^{\rm
inv}_\mu(a)=-ia\star\xi_\mu,\qquad\nabla^{\rm
inv}(X_{\mu\nu})(a)\equiv \nabla^{\rm
inv}_{(\mu\nu)}(a)=-ia\star(\xi_\mu\xi_\nu),\nonumber
\\
{\cal{A}}(\partial_\mu)\equiv{\cal{A}}_\mu=-i(A_\mu-\xi_\mu),\qquad
{\cal{A}}(X_{(\mu\nu)})\equiv{\cal{A}}_{(\mu\nu)}=-i(A_{(\mu\nu)}-\xi_\mu\xi_\nu),
\label{arondfinal}
\end{gather}
where the subscript $(\mu\nu)$ denotes symmetry under the exchange
of $\mu$ and $\nu$. Then, any NC connection is obtained as the sum
of the canonical connection and the tensor form, namely
\begin{gather*}
\nabla_\mu(a)=\nabla^{\rm inv}_\mu(a)+{\cal{A}}_\mu\star
a=\partial_\mu a-iA_\mu\star a,
\\
\nabla_{(\mu\nu)}(a)=\nabla^{\rm
inv}_{(\mu\nu)}(a)+{\cal{A}}_{(\mu\nu)}\star
a=[i\xi_\mu\xi_\nu,a]_\star-iA_{(\mu\nu)}\star a.
\end{gather*}

From this, one obtains the following expressions for the curvature
\begin{proposition}
Consider the differential calculus based on the maximal subalgebra
of derivations of the Moyal algebra related to symplectomorphisms.
The components of the $2$-form curvature of a NC connection
defined by a tensor $1$-form with components
${\cal{A}}_\mu,{\cal{A}}_{(\mu\nu)}$ are given by
\begin{gather}
F(\partial_\mu,\partial_\nu)\equiv
F_{\mu\nu}=[{\cal{A}}_\mu,{\cal{A}}_\nu]_\star-i\Theta^{-1}_{\mu\nu},
\label{curv1}
\\
F(\partial_\mu ,X_{(\rho\sigma)})\equiv
F_{\mu(\rho\sigma)}=[{\cal{A}}_\mu,{\cal{A}}_{(\rho\sigma)}]_\star-\Theta^{-1}_{\mu\rho}{\cal{A}}_\sigma
-\Theta^{-1}_{\mu\sigma}{\cal{A}}_\rho, \label{curv2}
\\
F(X_{(\mu\nu)},X_{(\rho\sigma)})\equiv
F_{(\mu\nu)(\rho\sigma)}=[{\cal{A}}_{(\mu\nu)},{\cal{A}}_{(\rho\sigma)}]_\star
+\Theta^{-1}_{\rho\nu}{\cal{A}}_{(\mu\sigma)} \nonumber\\
\qquad{}+
\Theta^{-1}_{\sigma\nu}{\cal{A}}_{(\mu\rho)}+\Theta^{-1}_{\rho\mu}{\cal{A}}_{(\nu\sigma)}
+\Theta^{-1}_{\sigma\mu}{\cal{A}}_{(\nu\rho)}.\label{curv3}
\end{gather}
\end{proposition}

\begin{proof}
Use ${F}^{\rm inv}{(X,Y)}\equiv\eta{[X,Y]}-[\eta(X),\eta(Y)]$ to
evaluate the curvature for the canonical connection. Consider
f\/irst $F^{\rm inv}_{\mu\nu}$. From linearity of $\eta$,
$[\partial_\mu,\partial_\nu]=0$ and
$[\eta_\mu,\eta_\nu]=i\Theta^{-1}_{\mu\nu}$, one f\/inds $F^{\rm
inv}_{\mu\nu}=-i\Theta^{-1}_{\mu\nu}$. Then, from
\eqref{eq:final3}, one gets~\eqref{curv1}. To obtain
\eqref{curv2}, compute
$[\partial_\mu,X_{(\rho\sigma)}](a)=[{\textup{Ad}}_{\eta_\mu},{\textup{Ad}}_{\eta_{(\rho\sigma)}}]
={\textup{Ad}}_{[\eta_\mu,\eta_{\rho\sigma}]_\star}$ using
\eqref{eq:addicom}. This yields
$[\partial_\mu,X_{(\rho\sigma)}](a)=\Theta^{-1}_{\mu\rho}\partial_\sigma
a+\Theta^{-1}_{\mu\sigma}\partial_\rho a$ so that
$\eta([\partial_\mu,X_{(\rho\sigma)}])=\Theta^{-1}_{\mu\rho}\eta_\sigma
+\Theta^{-1}_{\mu\sigma}\eta_\rho$which yields $F^{\rm
inv}_{\mu(\rho\sigma)}=0$. This combined with \eqref{eq:final3}
yields~\eqref{curv2}. For \eqref{curv3}, compute
$[X_{(\mu\nu)},X_{(\rho\sigma)}](a)={\textup{Ad}}_{[\eta_{(\mu\nu)},\eta_{(\rho\sigma)}]_\star}$
using \eqref{eq:slnr}. A straightforward calculation yields
$F^{\rm inv}_{(\mu\nu)(\rho\sigma)}=0$. From this
follows~\eqref{curv3}.
\end{proof}

The gauge transformations are
\begin{gather*}
A_\mu^g=g^\dag\star A_\mu\star g+ig^\dag\star\partial_\mu g,\\
 A_{(\mu\nu)}^g=g^\dag\star A_{(\mu\nu)}\star g+ig^\dag\star(\xi_\mu\partial_\nu+\xi_\nu\partial_\mu)g,\qquad \forall\, g\in{\cal{U}}({\cal{M}}),
\\
{\cal{A}}_\mu^g=g^\dag\star{\cal{A}}_\mu\star g,\qquad
{\cal{A}}_{(\mu\nu)}^g=g^\dag\star{\cal{A}}_{(\mu\nu)}\star
g,\qquad \forall\, g\in{\cal{U}}({\cal{M}}),
\\
F_{\mu\nu}^g=g^\dag\star F_{\mu\nu}\star g,\qquad
F_{(\mu\nu)(\rho\sigma)}^g=g^\dag\star
F_{(\mu\nu)(\rho\sigma)}\star g,\qquad \forall\,
g\in{\cal{U}}({\cal{M}}).
\end{gather*}
A possible construction of a NC gauge theory def\/ined from the
curvature \eqref{curv1}--\eqref{curv3} can be done as follows. Let
$[x]$ denote the mass dimension{\footnote{We work in the units
$\hbar=c=1$.}} of the quantity $x$. First, perform the rescaling
$\eta_{(\mu\nu)}\to\mu\theta\eta_{(\mu\nu)}$ where $\mu$ is a
parameter with $[\mu]=1$ that will f\/ix the mass scale of the
Higgs f\/ield to be identif\/ied in a while. Accordingly, the
commutation relations \eqref{eq:slnr}, \eqref{eq:addicom} are
modif\/ied as
\begin{gather*}
[\eta_{(\mu\nu)},\eta_{(\rho\sigma)}]_\star=-\mu\theta
\big(\Theta^{-1}_{\rho\nu}{\eta_{(\mu\sigma)}}+\Theta^{-1}_{\sigma\nu}{\eta_{(\mu\rho)}}
+\Theta^{-1}_{\rho\mu}{\eta_{(\nu\sigma)}}+\Theta^{-1}_{\sigma\mu}{\eta_{(\nu\rho)}}\big),
\nonumber
\\
[\eta_\mu,\eta_{(\rho\sigma)}]_\star=\mu\theta\big(\Theta^{-1}_{\mu\rho}\eta_\sigma
+\Theta^{-1}_{\mu\sigma}\eta_\rho\big),
\end{gather*}
and the components of the curvature become
\begin{gather*}
 F_{\mu(\rho\sigma)}=[{\cal{A}}_\mu,{\cal{A}}_{(\rho\sigma)}]_\star-\mu\theta\big(\Theta^{-1}_{\mu\rho}{\cal{A}}_\sigma
-\Theta^{-1}_{\mu\sigma}{\cal{A}}_\rho\big), 
\\ F_{(\mu\nu)(\rho\sigma)}=[{\cal{A}}_{(\mu\nu)},{\cal{A}}_{(\rho\sigma)}]_\star+\mu\theta\big(\Theta^{-1}_{\rho\nu}{\cal{A}}_{(\mu\sigma)} + \Theta^{-1}_{\sigma\nu}{\cal{A}}_{(\mu\rho)}+\Theta^{-1}_{\rho\mu}{\cal{A}}_{(\nu\sigma)} +\Theta^{-1}_{\sigma\mu}{\cal{A}}_{(\nu\rho)}\big)
\end{gather*}
with \eqref{curv1} unchanged. Next, introduce a dimensionfull
coupling constant $\alpha$ with mass dimension $[\alpha]=2-n$
($D=2n$). The functional ${\cal{U}}({\cal{M}})$-gauge invariant
action is then def\/ined by
\begin{equation}
S(A_\mu, {\cal{A}}_{(\mu\nu)})=-{{1}\over{\alpha^2}}\int
d^{2n}x\big(F_{\mu\nu}\star F_{\mu\nu}+F_{\mu(\rho\sigma)}\star
F_{\mu(\rho\sigma)}+ F_{(\mu\nu)(\rho\sigma)}\star
F_{(\mu\nu)(\rho\sigma)}\big)\label{eq:actionym}
\end{equation}
and is chosen to depend on the f\/ields $A_\mu$ and
${\cal{A}}_{\mu\nu}$. The mass dimensions are
$[A_\mu]=[{\cal{A}}_\mu]=[{\cal{A}}_{(\mu\nu)}]=1$.\par Several
remarks are now in order. First, the purely spatial part
\eqref{curv1} takes the expected form when expressed in term of
$A_\mu$ through the 1st relation \eqref{arondfinal}, namely one
obtains easily  $F_{\mu\nu}=-i(\partial_\mu A_\nu-\partial_\nu
A_\mu-i[A_\mu,A_\nu]_\star)$. \par Then, one observes that
$F_{\mu(\rho\sigma)}$ can be reexpressed as
\begin{gather*}
F_{\mu(\rho\sigma)}=D^A_\mu{\cal{A}}_{(\rho\sigma)}-\mu\theta(\Theta^{-1}_{\mu\rho}{\cal{A}}_\sigma
-\Theta^{-1}_{\mu\sigma}{\cal{A}}_\rho), \qquad
D^A_\mu{\cal{A}}_{(\rho\sigma)}\equiv\partial_\mu{\cal{A}}_{(\rho\sigma)}-i[A_\mu,{\cal{A}}_{(\rho\sigma)}]_\star
\end{gather*}
using \eqref{eq:innerxi} and \eqref{arondfinal}.
$D^A_\mu{\cal{A}}_{(\rho\sigma)}$ can be interpreted as a
``covariant derivative'' describing a NC analogue of the minimal
coupling to the covariant f\/ield ${\cal{A}}_{\mu\nu}$. Besides,
one has
\begin{gather}
-{{1}\over{\alpha}}\int d^{2n}x F_{\mu(\rho\sigma)}\star F_{\mu(\rho\sigma)}=-{{1}\over{\alpha}}\int d^{2n}x (D^A_\mu{\cal{A}}_{(\rho\sigma)})^2\nonumber\\
\phantom{-{{1}\over{\alpha}}\int d^{2n}x F_{\mu(\rho\sigma)}\star
F_{\mu(\rho\sigma)}=}{}-4\mu\theta(D^A_\mu{\cal{A}}_{(\rho\sigma)})\Theta^{-1}_{\mu\sigma}{\cal{A}}_\sigma
+(4n+2)\mu^2{\cal{A}}_\mu{\cal{A}}_\mu. \label{eq:2emeterm}
\end{gather}
In view of the last term in \eqref{eq:2emeterm}, the gauge
invariant action \eqref{eq:actionym} involves a mass term for the
gauge potential proportional to $\sim
{{(4n+2)\mu^2}\over{\alpha}}{{A}}_\mu{{A}}_\mu$. Therefore, bare
mass terms for $A_\mu$ can appear while preserving the gauge
invariant of the action. Notice that the translational invariance
of the action is broken by the term
$(4n+2)\mu^2{\cal{A}}_\mu{\cal{A}}_\mu$ in view of
${\cal{A}}_\mu=-i(A_\mu-\xi_\mu)$. Translational invariance is
obviously maintened whenever in the action the gauge covariant
curvature $F_{\mu(\rho\sigma)}$ is replaced by the gauge covariant
derivative $D^A_\mu{\cal{A}}_{(\rho\sigma)}$.

The fact that $D^A_\mu{\cal{A}}_{(\rho\sigma)}$ in
\eqref{eq:actionym} can be viewed as a NC analogue of the
covariant derivative of ${{D(D+1)}\over{2}}$ scalar f\/ields
${\cal{A}}_{\mu\nu}$, is very reminiscent of a Yang--Mills--Higgs
action for which the covariant coordinates ${\cal{A}}_{\mu\nu}$
play the role of Higgs f\/ields. Then, the last term in the action
\eqref{eq:actionym} which is the square of
$F_{(\mu\nu)(\rho\sigma)}$ can be interpreted as the Higgs
(quartic) potential part. Therefore, the use of a dif\/ferential
calculus based on the maximal subalgebra of
${\text{\textup{Der}}}({\cal{M}})$ whose elements generate
inf\/initesimal symplectomorphisms permits one to construct
naturally NC analogues of Yang--Mills--Higgs actions def\/ined on
Moyal space. Note that any attempt to interpret
${\cal{A}}_{\mu\nu}$ as possibly related to a gravitational
f\/ield, owing simply to the fact that it is a symmetric tensor,
would be physically misleading at least in view of the fact that
the symmetry group of the Moyal space is $SO(D)\cap SP(D)$ but not
$SO(D)$.

\section{Discussion}\label{section4}

The derivation-based dif\/ferential calculus is a mathematical
algebraic framework that permits one to generate from a given
associative algebra dif\/ferent consistent dif\/ferential calculi.
The case of Moyal algebras has been considered in the present
paper. Let us compare this latter situation with two other
noncommutative geometries, which share some common structures with
the one studied here.

First consider $M_n({\mathbb{C}})$, the f\/inite dimensional
algebra of $n\times n$ matrices. The algebra $M_n({\mathbb{C}})$
has only inner derivations, a trivial center ${\mathbb{C}}$ and
admits canonical NC gauge invariant connections. This last
property is insured by the existence in each case of a
${\mathbb{C}}$-linear map $\eta:
{\text{\textup{Der}}}({\mathbb{A}})  \rightarrow {\mathbb{A}}$
such that $X(a) = [\eta(X), a]$ for any $a \in {\mathbb{A}}$
(${\mathbb{A}}=M_n({\mathbb{C}})$ or ${\cal{M}}$). This map
def\/ines a canonical gauge invariant connection on (the right
${\mathbb{A}}$-module) ${\mathbb{A}}$: $a \mapsto \nabla_{X} a = -
a \eta(X)$. For the dif\/ferential calculus based on a subalgebra
of ${\text{\textup{Der}}}({\cal{M}})$, as considered here, the map
$\eta$ is def\/ined by~\eqref{eq:thetaform}. For the
dif\/ferential calculus based on
${\text{\textup{Der}}}(M_n({\mathbb{C}}))$, the map is def\/ined
by the canonical $1$-form connection $i\theta$ of
$M_n({\mathbb{C}})$,
$i\theta\in\Omega^1_{{\text{\textup{Der}}}}(M_n({\mathbb{C}}))$
interpreted as a map ${\textup{Int}}(M_n({\mathbb{C}}))\to
M_n({\mathbb{C}})$ and such that
$i\theta({\textup{Ad}}_\gamma)=\gamma-{{1}\over{n}}Tr(\gamma){\mathbb{I}}$,
$\forall\gamma\in M_n({\mathbb{C}})$. However, this last map is a
morphism of Lie algebra from
${\text{\textup{Der}}}(M_n({\mathbb{C}}))$ to the subalgebra of
$M_n({\mathbb{C}}$ formed by the traceless elements (the Lie
bracket is the usual commutator) and therefore, the curvature of
the canonical connection is zero. This is not the case for the
dif\/ferential calculus considered here. The map $\eta$ def\/ined
in~\eqref{eq:thetaform} is not a~Lie algebra morphism which is
signaled by a non zero curvature for the canonical connection.

Consider now the algebra ${\mathbb{A}}=C^\infty(M)\otimes
M_n({\mathbb{C}})$ of matrix valued functions on a smooth f\/inite
dimensional manifold $M$ whose derivation-based dif\/ferential
calculus was f\/irst considered in \cite{Dubois-Violette:1989vq}.
In the present case, ${\cal{Z}}({\mathbb{A}}) = C^\infty(M)$ and
${\text{\textup{Der}}}({\mathbb{A}}) =
[{\text{\textup{Der}}}(C^\infty(M))\otimes {\mathbf{1}} ] \oplus [
C^\infty(M) \otimes {\text{\textup{Der}}}(M_n) ] = \Gamma(M)
\oplus [C^\infty(M) \otimes {\mathfrak{sl}}_n]$ in the sense of
Lie algebras and $C^\infty(M)$-modules. $\Gamma(M)$ is the Lie
algebra of smooth vector f\/ields on $M$. Then, for any derivation
${\cal{X}}\in{\text{\textup{Der}}}({\mathbb{A}})$, ${\cal{X}} = X
+ ad_\gamma$ with $X \in \Gamma(M)$ and $\gamma \in C^\infty(M)
\otimes {\mathfrak{sl}}_n$, the traceless elements in
${\mathbb{A}}$. Set ${\mathbb{A}}_0=C^\infty(M) \otimes
{\mathfrak{sl}}_n$. One can identify ${\textup{Int}}({\mathbb{A}})
={\mathbb{A}}_0 $ and ${\text{\textup{Out}}}({\mathbb{A}}) =
\Gamma(M)$. Therefore, one has both inner and outer derivations.
Finally, one has $\Omega^\ast_{\text{\textup{Der}}}({\mathbb{A}})=
\Omega^\ast(M) \otimes \Omega^\ast_{\text{\textup{Der}}}(M_n)$
with a dif\/ferential ${\hat{\mathrm{d}}} = {\mathrm{d}} +
{\mathrm{d}}'$, where ${\mathrm{d}}$ is the de~Rham dif\/ferential
and ${\mathrm{d}}'$ is the dif\/ferential operating on the
$M_n({\mathbb{C}})$ part. The $1$-form related to the canonical
connection is def\/ined by $i\theta({\cal{X}})=\gamma$. As a map
from ${\text{\textup{Der}}}({\mathbb{A}})$ to ${\mathbb{A}}_0$, it
def\/ines a~splitting of Lie algebras and $C^\infty(M)$-modules of
the short exact sequence
\begin{gather*}
0\longrightarrow{\mathbb{A}}_0\longrightarrow{\text{\textup{Der}}}
({\mathbb{A}})\longrightarrow{\text{\textup{Out}}}({\mathbb{A}})\longrightarrow0
\end{gather*}
while the map $\eta$ def\/ined by \eqref{eq:thetaform} does not
have a similar property. The canonical connection on (the right
${\mathbb{A}}$-module) ${\mathbb{A}}$ is def\/ined from $-i\theta$
by
$\nabla_{\cal{X}}(a)={\cal{X}}(a)-i\theta({\cal{X}})a=X(a)-a\gamma$,
$\forall\, a\in{\mathbb{A}}$ but is not gauge invariant while the
corresponding curvature is zero, due to the above property of
splitting of Lie algebras. Past classical studies of the
corresponding gauge theories, with an action constructed mainly
from the square of the curvature, gave rise to the interpretation
of the gauge potential as involving two parts, one being the
``ordinary'' gauge theories and the other one identif\/iable as
Higgs f\/ields. Indeed, one can show that the simple action $\sim$
$\int d^Dx F_{\mu \nu} F^{\mu \nu}$ constructed using the
corresponding curvature $F_{\mu \nu}$ exhibits non trivial vacuum
states in the Higgs part, from which a mass generation on the
``ordinary'' gauge f\/ields is a consequence. This situation has
been generalised \cite{DBV-M, Masson:1999} to the case of the
algebra of endomorphisms of a $SU(n)$-vector bundle in the sense
that the situation of the trivial bundle correspond to the algebra
of matrix-valued functions. Because of the possible non trivial
global topology of the bundle, the situation is more involved
\cite{DBV-M} but reveals essentially that this physical
interpretation of the components of the noncommutative gauge
f\/ields can be performed in the same way.

The Yang--Mills--Higgs type action constructed from dif\/ferential
calculus based on the subalgebra
${\cal{G}}_2\subset{\text{\textup{Der}}}({\cal{M}})$ in
Section~\ref{section3} shares common features with this last
situation: Each additional inner derivation supplementing the
``ordinary spatial derivations'', which may be viewed as related
to an extra noncommutative dimension, corresponds to an additional
covariant coordinate that can be interpreted as a Higgs f\/ield.
Covariant coordinates have thus a natural interpretation as Higgs
f\/ields within the framework of the present derivation-based
dif\/ferential calculus. Then, Yang--Mills--Higgs models can be
obtained from actions built from the square of the curvature
\eqref{curv1}--\eqref{curv3}.

In these type of models, the polarisation tensor for the gauge
potential $A_\mu$ still exhibits an IR singularity similar to the
one given in \eqref{eq:singul1} with however a dif\/ferent overall
factor depending in particular on the dimension $D$ of the
noncommutative space and the Higgs f\/ields content. The
calculation can be performed by using auxiliary integrals given
below. We consider here the case where no bare mass term for the
gauge potential is present. Inclusion of a bare mass term would
not alter the conclusion. It is convenient to set
$\Phi_a\equiv{\cal{A}}_{\mu\nu}$, $a=1,\dots ,{{D(D+1)}\over{2}}$
and parametrise the gauge invariant action as (we def\/ine
$p\wedge k$$\equiv$$p_\mu\Theta_{\mu\nu}k_\nu$)
\begin{equation*}
S_{cl}=\int d^Dx{{1}\over{4}}\big(F_{\mu\nu}\star F_{\mu\nu}+(D_\mu\Phi_a)^2+F_{ab}\star F_{ab}\big) , 
\end{equation*}
where the coupling constant $\alpha$ has been set equal to $1$ and
$D_\mu\Phi_a$$=$$[{\cal{A}}_\mu,\Phi_a]_\star$$=$$\partial_\mu\Phi_a-i[A_\mu,\Phi_a]_\star$.
The action $S_{cl}$ must be supplemented by a BRST-invariant gauge
f\/ixing term $S_{GF}$, which can be taken as
\begin{gather*}
S_{GF}=s\int d^Dx\left({\bar{C}}\partial^\mu
A_\mu+{{\lambda}\over{2}}{\bar{C}}b \right)=\int
d^Dx\left(b\partial^\mu A_\mu-{\bar{C}}\partial^\mu(\partial_\mu
C-i[A_\mu,C]_\star)+{{\lambda}\over{2}}b^2\right),
\end{gather*}
where the nilpotent Slavnov operation $s$ is def\/ined through the
following BRST structure equations
\begin{gather*}
sA_\mu=\partial_\mu C-i[A_\mu,C]_\star,\qquad sC=iC\star C,\qquad
s{\bar{C}}=b,\qquad sb=0,
\end{gather*}
where $\lambda$ is a real constant and $C$, ${\bar{C}}$ and $b$
denote respectively the ghost, the antighost and the Stuekelberg
auxiliary f\/ield with ghost number respectively equal to $+1$,
$-1$ and $0$. Recall that $s$ acts as a graded derivation on the
various objects with grading def\/ined by the sum of the degree of
dif\/ferential forms and ghost number (modulo~2). In the
following, we will perform the calculation using a Feynman-type
gauge. Accordingly, the propagator for the $A_\mu$ in momentum
space takes the diagonal form
$G_{\mu\nu}(p)$$=$$\delta_{\mu\nu}/p^2$. The ghost and Higgs
propagators are respectively given by $G_g(p)$$=$$1/p^2$ and
$G^H_{ab}(p)$$=$$2\delta_{ab}/(p^2+\mu^2)$. The Feynman rules used
in the course of the calculation are given in the
Appendix~\ref{Feynman}.

\begin{figure}[t]
\centerline{\includegraphics[scale=0.4]{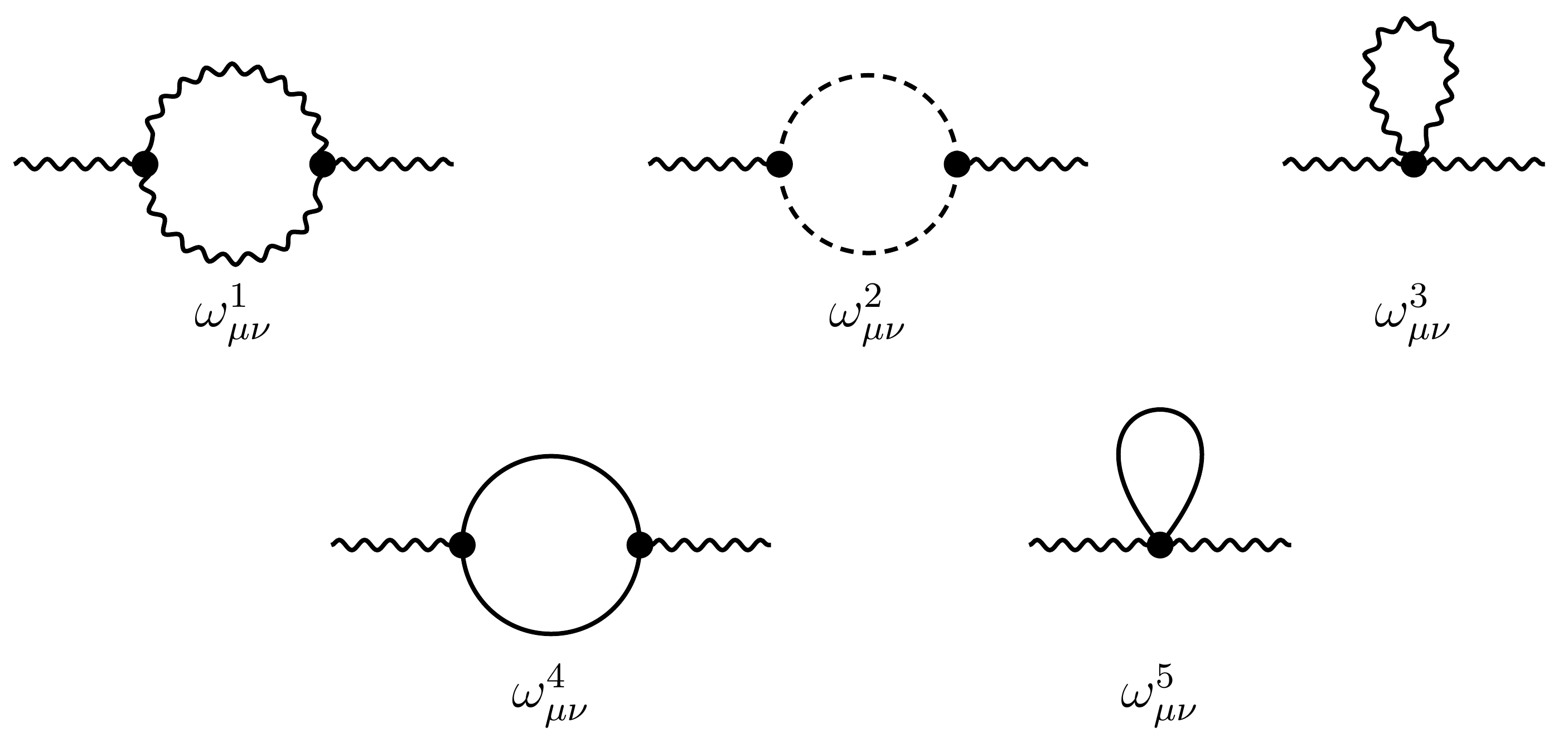}}
  \caption{One-loop diagrams contributing to the vacuum polarisation tensor. The wavy lines correspond to $A_\mu$.
The full (resp. dashed) lines correspond to the $\Phi$ (resp.
ghost) f\/ields.}
  \label{fig:figure1}
\end{figure}

The one-loop diagrams contributing to the vacuum polarisation
tensor $\omega_{\mu\nu}(p)$ are depicted on the
Fig.~\ref{fig:figure1}. The respective contributions can be
written as
\begin{gather}
\omega^1_{\mu\nu}(p)=4\int
{{d^Dk}\over{(2\pi)^D}}{{\sin^2({{p\wedge
k}\over{2}})}\over{k^2(p+k)^2}}\big[((k-p)^2+(k+2p)^2)\delta_{\mu\nu}+(D-6)p_\mu
p_\nu \nonumber
\\
\phantom{\omega^1_{\mu\nu}(p)=}{} +(p_\mu k_\nu+k_\mu
p_\nu)(2D-3)+k_\mu k_\nu(4D-6)\big],\label{eq:omega1}
\\
\omega^2_{\mu\nu}(p)=4\int {{d^Dk}\over{(2\pi)^D}}
{{\sin^2({{p\wedge k}\over{2}})}\over{k^2(p+k)^2}}k_\mu
k_\nu,\qquad \omega^3_{\mu\nu}(p)=8(D-1)\delta_{\mu\nu}\int
{{d^Dk}\over{(2\pi)^D}} {{\sin^2({{p\wedge
k}\over{2}})}\over{k^2}},\label{eq:omega23}
\\
\omega^4_{\mu\nu}(p)=4{\cal{N}}\int {{d^Dk}\over{(2\pi)^D}}
{{\sin^2({{p\wedge
k}\over{2}})}\over{(k^2+\mu^2)((p+k)^2+\mu^2)}}(p+2k)_\mu(p+2k)_\nu,
\label{eq:omega4}
\\
\omega^5_{\mu\nu}(p)=-4{\cal{N}}\delta_{\mu\nu}\int
{{d^Dk}\over{(2\pi)^D}} {{\sin^2({{p\wedge
k}\over{2}})}\over{(k^2+\mu^2)}}, \label{eq:omega5}
\end{gather}
where ${\cal{N}}$ is the number of $\Phi$ f\/ields, i.e.\
${\cal{N}}$$=$${{D(D+1)}\over{2}}$ for $sp(D,{\mathbb{R}})$.

It is convenient here to use the Feynman parametrisation, namely
${{1}\over{ab}} = \int_0^1 dx{{1}\over{(xa+(1-x)b)^2}}$. Then,
standard manipulations permit one to extract the IR limit of
\eqref{eq:omega1}--\eqref{eq:omega5}, denoted by $\omega^{\rm
IR}_{\mu\nu}(p)$. In the course of the derivation, it is very
convenient to use the following integrals
\begin{gather*}
J_N({\tilde{p}})\equiv\int
{{d^Dk}\over{(2\pi)^D}}{{e^{ik{\tilde{p}}}}\over{(k^2+m^2)^N}}=a_{N,D}{\cal{M}}_{N-{{D}\over{2}}}(m|{\tilde{p}}|),
\\
J_{N,\mu\nu}({\tilde{p}})\equiv\int {{d^Dk}\over{(2\pi)^D}}{{k_\mu k_\nu  e^{ik{\tilde{p}}}}\over{(k^2+m^2)^N}}=a_{N,D}\big(\delta_{\mu\nu}{\cal{M}}_{N-1-{{D}\over{2}}}(m|{\tilde{p}}|)-{\tilde{p}}_\mu{\tilde{p}}_\nu{\cal{M}}_{N-2-{{D}\over{2}}}(m|{\tilde{p}}|) \big), 
\end{gather*}
where
\begin{gather*}
a_{N,D}={{2^{-({{D}\over{2}}+N-1)}}\over{\Gamma(N)\pi^{{D}\over{2}}}},\qquad
{\cal{M}}_Q(m|{\tilde{p}}|)={{1}\over{(m^2)^Q}}(m|{\tilde{p}}|)^Q{\bf{K}}_Q(m|{\tilde{p}}|)
\end{gather*}
in which ${\bf{K}}_Q(z)$ is the modif\/ied Bessel function of
second kind and of order $Q \in {\mathbb{Z}}$ (recall
${\bf{K}}_{-Q}(z) = {\bf{K}}_Q(z)$) together with the asymptotic
expansion
\begin{gather*}
{\cal{M}}_{-Q}(m|{\tilde{p}}|)\sim2^{Q-1}{{\Gamma(Q)}\over{{\tilde{p}}^{2Q}}},\qquad
Q>0.
\end{gather*}
The IR limit of the vacuum polarisation tensor is given by
\begin{gather*}
\omega^{\rm IR}_{\mu\nu}(p)=(D+{\cal{N}}-2)\Gamma({{D}\over{2}})
{{{\tilde{p}}_\mu{\tilde{p}}_\nu}\over{{\pi^{D/2}(\tilde{p}}^2)^{D/2}}}+\cdots,
\end{gather*}
where the ellipses denote subleading singular terms. The overall
factor af\/fecting $\omega_{\mu\nu}^{\rm IR}(p)$ is modif\/ied
compared to \eqref{eq:singul1}. It cannot be canceled by tuning
the values for $D$ and ${\cal{N}}$. A~similar comment applies to
$\Pi_{ab}^{\rm IR}(p)$. Note that a one-loop calculation of the
polarisation tensor performed within some $N=1$ supersymmetric
version of the NC Yang--Mills theory~\cite{susy1} (see
also~\cite{susy2}) suggests a better IR behaviour due basically to
some compensation occurring between bosonic and fermionic loop
contributions.

\section[Graded differential calculus and gauge theories]{Graded dif\/ferential calculus and gauge theories}\label{section5}

In this section, we will construct a gauge theory starting from a
${\mathbb{Z}}_2$-graded version of the Moyal algebra as a direct
application of the framework for the ${\mathbb{Z}}_2$-graded
dif\/ferential calculus introduced in
Section~\ref{gradedcalc-section}.

Consider the ${\mathbb{Z}}_2$-graded vector space
${\mathbb{A}}^\bullet={\cal{M}}^0\oplus{\cal{M}}^1$ built from 2
copies of the Moyal algebra. ${\cal{M}}^0$ (resp.~${\cal{M}}^1$)
involves elements of homogeneous degree $0$ (resp.~$1$). We denote
any element $a\in{\mathbb{A}}^\bullet$ by $a=(a_0,a_1)$,
$a_i\in{\cal{M}}^i$, $i=1,2$.

\begin{proposition}
${\mathbb{A}}^\bullet={\cal{M}}^0\oplus{\cal{M}}^1$ equipped with
the internal product
\begin{gather*}
ab=(a_0\star b_0+a_1\star b_1,a_0\star b_1+a_1\star b_0), \qquad \forall \, a=(a_0,a_1), b=(b_0,b_1)\in{\cal{A}}^\bullet, 
\end{gather*}
where $\star$ is the Moyal product and the involution
$a^\dag=(a_0^\dag,ia_1^\dag)$ where $a_i^\dag$, $i=0,1$ is the
natural involution on the Moyal algebra, is an associative
$*$-algebra wit unit ${\mathbb{I}}=(1,0)$. The center is
${\cal{Z}}({\mathbb{A}}^\bullet)={\mathbb{C}}\oplus 0$.
\end{proposition}

\begin{proof}
Each part of the proposition can be proved by simple calculation.
Note that the determination of the center stems from
${\cal{Z}}({\cal{M}})={\mathbb{C}}$.
\end{proof}

An interesting generalisation of the NC gauge theory constructed
thorough Sections~\ref{map-eta} and~\ref{moyalcase} to a NC gauge
theory def\/ined on the ${\mathbb{Z}}_2$-graded algebra
${\cal{A}}^\bullet$ can be obtained as follows. We def\/ine
\begin{gather*}
T_\mu=(\eta_\mu,0),\qquad U_\mu=(0,\eta_\mu),\qquad M_{\mu\nu}=(\eta_{\mu\nu},0),\\
 J=(0,i),\qquad \eta_\mu=i\xi_\mu, \qquad \eta_{\mu\nu}=i\xi_\mu\xi_\nu,\qquad \mu,\nu=1,\dots ,D.
\end{gather*}
A useful relation can be easily shown
\begin{gather*}
[a,b]_\bullet=[(a_0,a_1),(b_0,b_1)]_\bullet=([a_0,b_0]_\star+\{a_1,b_1\}_\star,[a_0,b_1]_\star+[a_1,b_0]_\star),\qquad
\forall \,a,b\in{\mathbb{A}}^\bullet.
\end{gather*}
First, consider the Lie subalgebra of
${\text{\textup{Der}}}({\mathbb{A}}^\bullet)$ generated by the
real derivations of degree $0$ ${\textup{Ad}}_{T_\mu}$ and
${\textup{Ad}}_{M_{\mu\nu}}$, isomorphic to
${\cal{G}}_2\subset{\text{\textup{Der}}}({\cal{M}}^0)$ def\/ined
in Section~\ref{map-eta}. Then, enlarge this subalgebra by adding
the real derivation of degree $1$ ${\textup{Ad}}_{J}$ and
${\textup{Ad}}_{U_\mu}$. Notice that $J$ does not belong
to~${\cal{Z}}({\mathbb{A}}^\bullet)$ so that ${\textup{Ad}}_{J}$
is not the trivial $0$ derivation. One has the following
commutation relations
\begin{gather}
[T_\mu,T_\nu]_\bullet=i\Theta^{-1}_{\mu\nu}{\mathbb{I}},\qquad
[M_{\mu\nu},T_\rho]_\bullet=\big(\Theta^{-1}_{\nu\rho}T_\mu+\Theta^{-1}_{\mu\rho}T_\nu\big),\label{eq:prem-commut}
\\
[M_{\mu\nu},M_{\rho\sigma}]_\bullet=\big(\Theta^{-1}_{\nu\sigma}M_{\mu\rho}+\Theta^{-1}_{\nu\rho}M_{\mu\sigma}+
\Theta^{-1}_{\mu\sigma}M_{\nu\rho}+\Theta^{-1}_{\mu\rho}M_{\nu\sigma}\big),
\\
[U_\mu,U_\nu]_\bullet=i2M_{\mu\nu},\qquad
[T_\mu,U_\nu]_\bullet=\Theta^{-1}_{\mu\nu}J,\qquad
[M_{\mu\nu},U_\rho]_\bullet=\big(\Theta^{-1}_{\nu\rho}U_\mu+\Theta^{-1}_{\mu\rho}U_\nu\big),
\\
[J,J]_\bullet=-2{\mathbb{I}},\qquad [T_\mu,J]_\bullet=0,\qquad
 [M_{\mu\nu},J]_\bullet=0,\qquad [U_\mu,J]_\bullet=i2T_\mu . \label{eq:dern-commut}
\end{gather}

\begin{proposition}\label{g2dot}
The set of real derivations $\{{\textup{Ad}}_{a}\}$, $a=T_\mu,
U_\mu,M_{\mu\nu},J$ is a ${\mathbb{Z}}_2$-graded Lie subalgebra of
${\text{\textup{Der}}}({\mathbb{A}}^\bullet)$ and module over
${\cal{Z}}({\mathbb{A}}^\bullet)$.
\end{proposition}

\begin{proof}
The proposition is a direct consequence of the usual relation
$[{\textup{Ad}}_X,{\textup{Ad}}_Y]_\bullet={\textup{Ad}}_{[X,Y]_\bullet}$
combined with equations
\eqref{eq:prem-commut}--\eqref{eq:dern-commut}. The properties of
module are obvious.
\end{proof}

The map $\eta$ def\/ined in Section~\ref{map-eta} can be extended
as follows. One considers ${\cal{G}}_2^\bullet$, the image by the
map ${\textup{Ad}}$ of
$({\cal{P}}_2^0\oplus{\cal{P}}_1^1)\subset({\cal{M}}^0\oplus{\cal{M}}^1)$
where ${\cal{P}}_2^0\subset{\cal{M}}^0$ (resp.
${\cal{P}}_1^1\subset{\cal{M}}^1$) is the set of polynomial with
degree $d\le2$ (resp. $d\le1$). Then, the linear map $\eta$ of
Section~\ref{section3} can be extended to linear map of
homogeneous degree $0$
$\eta:{\cal{G}}_2^\bullet\to{\mathbb{A}}^\bullet\, /\,
\eta(X)=\eta({\textup{Ad}}_{(P_0,P_1)}=(P_0-P_0(0),P_1)$,
$P_0\in({\cal{P}}_2^0$, $P_1\in{\cal{P}}_1^1$.

We can now apply the algebraic scheme presented in
Section~\ref{gradedcalc-section}. The components of the $1$-form
$\eta$ are
\begin{gather*}
\eta({\textup{Ad}}_{T_\mu})=(\eta_\mu,0),\qquad
\eta({\textup{Ad}}_{U_\mu})=(0,\eta_\mu),\qquad
\eta({\textup{Ad}}_{M_{\mu\nu}})=(\eta_{\mu\nu},0),\qquad
\eta({\textup{Ad}}_J)=(0,i).
\end{gather*}
For any $a=(a_0,a_1)\in{\mathbb{A}}^\bullet, $the $1$-form
canonical connection on ${\mathbb{A}}^\bullet$ is determined by
\begin{gather*}
\nabla^{\rm
inv}_{{\textup{Ad}}_{T_\mu}}(a_0,0)=-(a_0\star\eta_\mu,0),\qquad
\nabla^{\rm
inv}_{{\textup{Ad}}_{T_\mu}}(0,a_1)=-(0,a_1\star\eta_\mu),
\\
\nabla^{\rm
inv}_{{\textup{Ad}}_{U_\mu}}(a_0,0)=-(0,a_0\star\eta_\mu),\qquad
\nabla^{\rm
inv}_{{\textup{Ad}}_{T_\mu}}(0,a_1)=(a_1\star\eta_\mu,0),
\\
\nabla^{\rm
inv}_{{\textup{Ad}}_{M_{\mu\nu}}}(a_0,0)=-(a_0\star\eta_{\mu\nu},0),\qquad
\nabla^{\rm
inv}_{{\textup{Ad}}_{M_{\mu\nu}}}(0,a_1)=-(0,a_1\star\eta_{\mu\nu}),
\\
\nabla^{\rm inv}_{{\textup{Ad}}_{J}}((a_0,0))=-(0,ia_0),\qquad
\nabla^{\rm inv}_{{\textup{Ad}}_{J}}(0,a_1)=(ia_1,0).
\end{gather*}
We now perform the rescaling $A(X)\to -iA(X)$ as in
Section~\ref{section3}. Then, for any NC connection on
${\mathbb{A}}^\bullet$ determined by the $1$-form
$A\in\Omega_{\text{\textup{Der}}}^{1,0}({\mathbb{A}}^\bullet)$,
the components of the tensor $1$-form are given~by
\begin{gather*}
{\cal{A}}({\textup{Ad}}_{T_\mu})=-i(A^0_\mu-\xi_\mu,0)\equiv-i({\cal{A}}^0_\mu,0),
\\
{\cal{A}}({\textup{Ad}}_{U_\mu})=-i(0,
A^1_\mu-\xi_\mu)\equiv-i(0,{\cal{A}}^1_\mu),
\\
{\cal{A}}({\textup{Ad}}_{M_{\mu\nu}})=-i(G^0_{\mu\nu}-\xi_\mu\xi_\nu,0)\equiv-i({\cal{G}}^0_{\mu\nu},0),
\\
{\cal{A}}({\textup{Ad}}_{J})=-i(0, \varphi-1)\equiv-i(0,\Phi),
\end{gather*}
where we have def\/ined $A({\textup{Ad}}_{T_\mu})\equiv
(A^0_\mu,0)$, $A({\textup{Ad}}_{U_\mu})\equiv (0,A^1_\mu)$,
$A({\textup{Ad}}_{M{_\mu\nu}})\equiv (G^0_{\mu\nu},0)$ and
$A({\textup{Ad}}_{J})\equiv (0,\varphi)$.

\begin{proposition}
Consider the restricted differential calculus based on
${\cal{G}}^\bullet_2\subset{\text{\textup{Der}}}({\mathbb{A}}^\bullet)$
given in Proposition~{\rm \ref{g2dot}}. The components of the
$2$-form curvature for a NC connection on ${\mathbb{A}}^\bullet$
are:
\begin{gather}
F({\textup{Ad}}_{T_\mu},{\textup{Ad}}_{T_\nu})=(-[{\cal{A}}^0_\mu,{\cal{A}}^0_\nu]_\star-i\Theta^{-1}_{\mu\nu},0),
\label{eq:courbe-1}\\
F({\textup{Ad}}_{U_\mu},{\textup{Ad}}_{U_\nu})=(-\{{\cal{A}}^1_\mu,{\cal{A}}^1_\nu\}_\star-2{\cal{G}}^0_{\mu\nu},0),
\label{eq:courbe-1+}
\\
F({\textup{Ad}}_J,{\textup{Ad}}_J)=(-2\Phi\star\Phi+2,0),\qquad
F({\textup{Ad}}_{T_\mu},{\textup{Ad}}_J)=(0,-[{\cal{A}}^0_\mu,\Phi]_\star),
\\
F({\textup{Ad}}_{U_\mu},{\textup{Ad}}_J)=(-\{{\cal{A}}^1_\mu,\Phi\}_\star-2{\cal{A}}^0_\mu,0),\qquad
F({\textup{Ad}}_{M_{\mu\nu}},{\textup{Ad}}_J)=(0,-[{\cal{G}}^0_{\mu\nu},\Phi]_\star),
\\
F({\textup{Ad}}_{T_\mu},{\textup{Ad}}_{U_\nu})=(0,-[{\cal{A}}^0_\mu,{\cal{A}}^1_\nu]_\star+i\Theta^{-1}_{\mu\nu}\Phi),
\\
F({\textup{Ad}}_{M_{\mu\nu}},{\textup{Ad}}_{T_\rho})
=(-[{\cal{G}}^0_{\mu\nu},{\cal{A}}^0_\rho]_\star+i\Theta^{-1}_{\nu\rho}{\cal{A}}^0_\mu
+i\Theta^{-1}_{\mu\rho}{\cal{A}}^0_\nu,0),
\\
F({\textup{Ad}}_{M_{\mu\nu}},{\textup{Ad}}_{U_\rho})=(0,-[{\cal{G}}^0_{\mu\nu},
{\cal{A}}^1_\rho]_\star+i\Theta^{-1}_{\nu\rho}{\cal{A}}^1_\mu+i\Theta^{-1}_{\mu\rho}{\cal{A}}^1_\nu),
\\
F({\textup{Ad}}_{M_{\mu\nu}},{\textup{Ad}}_{M_{\rho\sigma}})
=(-[{\cal{G}}^0_{\mu\nu},{\cal{G}}^0_{\rho\sigma}]_\star+
i(\Theta^{-1}_{\nu\sigma}{\cal{G}}^0_{\mu\rho}+\Theta^{-1}_{\nu\rho}{\cal{G}}^0_{\mu\sigma}
+\Theta^{-1}_{\mu\sigma}{\cal{G}}^0_{\nu\rho}+\Theta^{-1}_{\mu\rho}{\cal{G}}^0_{\nu\sigma}),0).\label{courbe-fin}
\end{gather}
\end{proposition}

\begin{proof}
The proposition can obtained by direct calculations.
\end{proof}

The gauge transformations can be obtained from
Proposition~\ref{graded-gauge}. After performing the rescaling
$A(X)\to-iA(X)$, one has for any
$g=(g_0,0)\in{\cal{U}}({\mathbb{A}}^\bullet)$
\begin{equation}
A^g({\textup{Ad}}_J)=ig^\dag{\textup{Ad}}_J(g)+g^\dag
A({\textup{Ad}}_J)g=g^\dag A({\textup{Ad}}_J)g,\label{gaugephi}
\end{equation}
where the second equality stems from $g^\dag{\textup{Ad}}_J(g)=0$
which holds since $g$ has degree $0$, which therefore transforms
as a tensor form despite the fact this quantity is actually
related to a $1$-form connection. From \eqref{gaugephi}, one
obtains
\begin{gather*}
\varphi^g=g_0^\dag\star\varphi\star g_0.
\end{gather*}
The other gauge transformations have standard expressions, in
particular the gauge transformations of all other components of
the $1$-form connection involve an inhomogeneous term.

Set now $A^0_\mu=A^1_\mu$. A gauge invariant action of the form
${\textup{Tr}}(|F(X,Y)|^2)$ by using
\eqref{eq:courbe-1}--\eqref{courbe-fin} where we have def\/ined
${\textup{Tr}}(a)=\int dx^{2n}a_0$, for any
$a=(a_0,a_1)\in{\mathbb{A}}^\bullet$ and $a_0\in{\cal{M}}$ such
that the integral exists. It is instructive to consider the case
where ${\cal{G}}^0_{\mu\nu}=0$ and to express the remaining
components of the curvature in terms of the components of the
$1$-form connection. We set again $F_{\mu\nu}=\partial_\mu
A_\nu-\partial_\nu A_\mu-i[A_\mu,A_\nu]_\star$. One obtains
\begin{gather*}
F({\textup{Ad}}_{T_\mu},{\textup{Ad}}_{T_\nu})=(-iF_{\mu\nu},0),\qquad F({\textup{Ad}}_{U_\mu},{\textup{Ad}}_{U_\nu})=(-\{A_\mu-\xi_\mu,A_\nu-\xi_\nu\}_\star,0),\\  
F({\textup{Ad}}_J,{\textup{Ad}}_J)=(-2\varphi\star\varphi+4\varphi,0),\qquad
F({\textup{Ad}}_{T_\mu},{\textup{Ad}}_J)=(0,-i(\partial_\mu\varphi-i[A_\mu,\varphi]_\star)),
\\
F({\textup{Ad}}_{U_\mu},{\textup{Ad}}_J)=(-\{A_\mu,\varphi\}_\star+2\xi_\mu\varphi,0),\qquad
F({\textup{Ad}}_{T_\mu},{\textup{Ad}}_{U_\nu})=(0,i(\Theta^{-1}_{\mu\nu}\varphi-F_{\mu\nu})).
\end{gather*}
To obtain the correct mass dimensions for the f\/ields and
parameters involved in the action, one performs, as in
Section~\ref{section3}, the rescaling $J\to{{1}\over{m\theta}}$,
$\eta_{\mu\nu}\to\mu\theta\eta_{\mu\nu}$ where the mass dimensions
of $\mu$ and $m$ are $[m]=[\mu]=1$. The mass dimensions for the
f\/ields are $[A_\mu]=[\varphi]=[G_{\mu\nu}]=1$.
$F({\textup{Ad}}_J,{\textup{Ad}}_J)$ is modif\/ied as
\begin{gather*}
F({\textup{Ad}}_J,{\textup{Ad}}_J)=\left(-2\varphi\star\varphi+{{4}\over{m\theta}}\varphi,0\right)
\end{gather*}
while the other are unchanged. Then, the gauge invariant action is
\begin{gather}
S(A_\mu,\varphi;{\cal{G}}_{\mu\nu}=0)={{1}\over{\alpha^2}}\int
d^{2n}x
(F_{\mu\nu}^2+\{{\cal{A}}_\mu,{\cal{A}}_\nu\}^2)+(\Theta^{-1}_{\mu\nu}\varphi-F_{\mu\nu})^2
\label{decadix-77}
\\
{}+(\partial_\mu\varphi-i[A_\mu,\varphi]_\star)^2+(\{A_\mu,\varphi\}_\star-2\xi_\mu\varphi)^2+
\left(\!4\varphi\star\varphi\star\varphi\star\varphi-{{8}\over{m\theta}}\varphi\star\varphi\star\varphi
+{{16}\over{m^2\theta^2}}\varphi^2\!\right).\nonumber
\end{gather}
The f\/irst two terms between parenthesis give rise to an action
of the form \eqref{eq:decadix1} of~\cite{de Goursac:2007gq,
Grosse:2007qx} which has been proposed as the gauge counterpart of
the renormalisable NC $\varphi^4$ model with harmonic
term~\cite{Grosse:2004yu}. However, this action is supplemented by
a Slavnov term \cite{slavnov} of the form $\sim\int
d^{2n}x\Theta^{-1}_{\mu\nu}\varphi F_{\mu\nu}$ which is
reminiscent of the so called $BF$ term (see e.g.~\cite{thomson,
superbf}). The last 3 terms in \eqref{decadix-77} describe an
action for the renormalisable NC $\varphi^4$ model with harmonic
term (supplemented however with a $\varphi^{\star3}$ potential
term) coupled in a gauge invariant way to $A_\mu$. The addition of
a Slavnov term to the simplest NC analogue of Yang--Mills action
$\int d^{2n}xF_{\mu\nu}\star F_{\mu\nu}$ is know to lead to an
action with a somewhat improved IR behaviour but does not prevent
the UV/IR mixing to occur in the gauge-f\/ixed action. Notice that
in the present situation~\eqref{decadix-77}, the f\/ield $\varphi$
propagates while in~\cite{slavnov} $\varphi$ serves as a simple
multiplier enforcing a zero-curvature constraint. The fact that
the NC $\varphi^4$ model with harmonic term appearing as a part of
a gauge action $\sim{\textup{Tr}}(|F(X,Y)|^2)$ is a consequence of
our choice for the ${\mathbb{Z}}_2$-graded algebra as well as of
the particular algebra of graded derivations generating the
dif\/ferential calculus. However, encoding both the gauge action
\eqref{eq:decadix1} and the NC $\varphi^4$ harmonic model within a
single gauge action ${\textup{Tr}}(|F(X,Y)|^2)$, and in particular
taking advantage of the explicit gauge invariance, may prove
useful to understand new features of the Langmann--Szabo duality
which plays an important role in the renormalisability of this NC
$\varphi^4$ and of its actual gauge counterpart (if any). To do
this, one has to consider the action with ${\cal{G}}_{\mu\nu}\ne0$
which can be viewed again as a Yang--Mills--Higgs action with
${\cal{G}}_{\mu\nu}$ possibly interpretable as a Higgs f\/ield.
Note f\/inally that starting with actions
$\sim{\textup{Tr}}(|F(X,Y)|^2)$ permits one to deal with simple
vacua.

\section{Summary}\label{section6}

This section summarises the main features and results of the
discussion. First, a general way to construct gauge models from
the algebraic scheme presented in Section~\ref{section2} is as
follows. Consider $\mathbb{A}$ as a module over itself and pick
some Lie subalgebra of derivations of $\mathbb{A}$,
${\cal{G}}\subset \textup{Der}(\mathbb{A})$, which is also a
module over the center of $\mathbb{A}$. Then, a dif\/ferential
calculus can be def\/ined from
Propositions~\ref{prop2},~\ref{prop3} or whenever $\mathbb{A}$ is
${\mathbb{Z}}_2$-graded from Proposition~\ref{gradedcalculus}.
Correspondingly, NC connections and curvatures can then be
def\/ined as given in Def\/inition~\ref{def2} and
Proposition~\ref{prop5} in the non-graded case or, when
$\mathbb{A}$ is ${\mathbb{Z}}_2$-graded, from
Def\/inition~\ref{connection-curv}. Then, gauge actions can be
obtained by considering the ``square'' of the curvatures. This
general construction is applied f\/irst to the Moyal algebra, as
discussed in Sections~\ref{section3} and~\ref{section4}, and to a
$\mathbb{Z}_2$-graded associative algebra built from two copies of
the Moyal algebra.

These two dif\/ferent situations have common features. In each
case, the essential properties of the gauge models stem from the
existence of a $\mathbb{C}$-linear map
$\eta:{\cal{G}}\to\mathbb{A}$. Then, since all the considered
${\cal{G}}$'s involve only inner derivations,
Proposition~\ref{prop2.9} (or Lemma~\ref{lemma} in the
$\mathbb{Z}_2$-graded case) ensures the existence of a canonical
connection which turns out to be invariant under the gauge
transformations. From this follows a natural construction of
tensor forms, i.e.\ the so-called covariant coordinates, that will
play the role of Higgs f\/ields in the gauge actions. More
physically, the additional derivations supplementing the ordinary
``spatial derivations'' that are involved in the Lie subalgebras
of derivations considered below may be interpreted as related to
NC directions associated with the Higgs f\/ields.

In the non graded case presented in Sections~\ref{section3}
and~\ref{section4}, $\mathbb{A}={\cal{M}}$ and we consider the
dif\/ferential calculus based on the maximal Lie subalgebra
${\cal{G}}_2\subset\textup{Der}({\cal{M}})$ that is related to
symplectomorphisms. The resulting gauge models stemming from the
above scheme can be interpreted as Yang--Mills--Higgs models on
Moyal spaces, which however still suf\/fer apparently from UV/IR
mixing. These models have, in some sense, some similarity with the
NC gauge models based on  ${\mathbb{A}}=C^\infty(M)\otimes
M_n({\mathbb{C}})$ introduced in \cite{Dubois-Violette:1989vq}.

In the $\mathbb{Z}_2$-graded case where $\mathbb{A}$ is built from
2 copies of the Moyal algebra that is considered in
Section~\ref{section5}, we consider a dif\/ferential calculus
based on a Lie subalgebra of $\textup{Der}(\mathbb{A})$ that may
be viewed as a natural extension of ${\cal{G}}_2$. The resulting
gauge models can again be interpreted as Yang--Mills--Higgs type
models on this graded associative algebra. The gauge action
involves now three interesting contributions. The f\/irst one is
the action as given in \eqref{eq:decadix1} which has been proposed
in~\cite{de Goursac:2007gq} and \cite{Grosse:2007qx}  as a gauge
counterpart of the renormalisable NC $\varphi^4$ model with
harmonic term. The second contribution is a Slavnov term as the
one initially introduced in~\cite{slavnov} as a f\/irst attempt to
cure the UV/IR mixing in the naive gauge theory on Moyal spaces.
Its actual ef\/fect on the UV/IR mixing on the gauge action
def\/ined on ${\mathbb{A}}^\bullet={\cal{M}}^0\oplus{\cal{M}}^1$
that we have obtained in Section~\ref{section5} remains to be
analysed. Notice that in the present situation, the vacuum
conf\/iguration for the action is trivial so that the
dif\/f\/icult problem to deal with a~complicated vacuum (see
\cite{de Goursac:2007gq,vacuumym:2008}) is absent here. Finally,
the full gauge action involves, as the third interesting
contribution, the NC renormalisable $\varphi^4$ model with
harmonic term (up to an additional $\phi^3$ interaction term),
therefore exhibiting a link between this latter renormalisable NC
scalar theory and gauge theories. In this framework, notice that
the harmonic term may be interpreted as a~quartic gauge potential
coupling, built from the (gauge potential of the) canonical
connection and the component of the gauge connection corresponding
to the derivation generated by $(0,i)$ (see
Section~\ref{section5}).

A next step would be to study whether the vacuum polarisation
tensor still involves an IR singularity of the type given by
\eqref{eq:singul1} which would then probe the actual ef\/fect of
anticommutator term $\sim\{{\cal{A}}_\mu,{\cal{A}}_\nu\}^2_\star$
(combined to a Slavnov term) on the UV/IR mixing. This amounts to
consider the full action and perform a gauge f\/ixing, e.g.\ using
the BRST formalism.

\appendix
\section{Relevant Feynman rules}
\label{Feynman}

In the following vertex functions, momentum conservation is
understood. All the momenta are incoming. We def\/ine $p\wedge
k$$\equiv$$p_\mu\Theta_{\mu\nu}k_\nu$.
\begin{itemize}\itemsep=0pt
\item 3-gauge boson vertex:
\begin{gather*}
V^3_{\alpha\beta\gamma}(k_1,k_2,k_3)=-i2\sin\!\left({{k_1\wedge
k_2}\over{2}}\right)\!\big[(k_2-k_1)_\gamma\delta_{\alpha\beta}+(k_1-k_3)_\beta
\delta_{\alpha\gamma}+(k_3-k_2)_\alpha\delta_{\beta\gamma}\big];
\end{gather*}
\item 4-gauge boson vertex:
\begin{gather*}
V^4_{\alpha\beta\gamma\delta}(k_1,k_2,k_3,k_4)=-4\left[
(\delta_{\alpha\gamma}\delta_{\beta\delta}-\delta_{\alpha\delta}\delta_{\beta\gamma})\sin\left({{k_1\wedge
k_2}\over{2}}\right)\sin\left({{k_3\wedge
k_4}\over{2}}\right)\right.\nonumber
\\
\phantom{V^4_{\alpha\beta\gamma\delta}(k_1,k_2,k_3,k_4)=}{}+(\delta_{\alpha\beta}\delta_{\gamma\delta}-\delta_{\alpha\gamma}\delta_{\beta\delta})\sin\left({{k_1\wedge
k_4}\over{2}}\right)\sin\left({{k_2\wedge k_3}\over{2}}\right) \\
\left.\phantom{V^4_{\alpha\beta\gamma\delta}(k_1,k_2,k_3,k_4)=}{}
+(\delta_{\alpha\delta}\delta_{\beta\gamma}-\delta_{\alpha\beta}\delta_{\gamma\delta})\sin\left({{k_3\wedge
k_1}\over{2}}\right)\sin({{k_2\wedge k_4}\over{2}})\right];
\end{gather*}
\item gauge boson-ghost $V^g_{\mu}(k_1,k_2,k_3)$ and gauge
boson-Higgs $V^H_{ab\mu}(k_1,k_2,k_3)$ vertices:
\begin{gather*}
V^g_{\mu}(k_1,k_2,k_3)=i2k_{1\mu}\sin\left({{k_2\wedge
k_3}\over{2}}\right),\\
V^H_{ab\mu}(k_1,k_2,k_3)=i\delta_{ab}(k_1-k_2)_\mu\sin\left({{k_2\wedge
k_3}\over{2}}\right);
\end{gather*}
\item Seagull vertex:
\begin{gather*}
V^s_{ab\alpha\beta}(k_1,k_2,k_3,k_4)=-2\delta_{\alpha\beta}\delta_{ab}\left[\cos\left({{k_3\wedge k_1+k_4\wedge k_2}\over{2}}\right)\right.\\
\left.\phantom{V^s_{ab\alpha\beta}(k_1,k_2,k_3,k_4)=}{} -
\cos\left({{k_1\wedge k_2}\over{2}}\right)\cos\left({{k_3\wedge
k_4}\over{2}}\right) \right];
\end{gather*}
\item   3-Higgs $V^H_{abc}(k_1,k_2,k_3)$ vertex
\begin{gather*}
V^H_{abc}(k_1,k_2,k_3)=iC_{ab}^c\sin\left({{k_1\wedge
k_2}\over{2}}\right);
\end{gather*}
\item 4-Higgs vertex:
\begin{gather*}
V^H_{abcd}(k_1,k_2,k_3,k_4)=4\left[
(\delta_{ac}\delta_{bd}-\delta_{ad}\delta_{bc})\sin\left({{k_1\wedge
k_2}\over{2}}\right)\sin\left({{k_3\wedge k_4}\over{2}}\right)
\right.\nonumber
\\
\phantom{V^H_{abcd}(k_1,k_2,k_3,k_4)=}{}
+(\delta_{ab}\delta_{cd}-\delta_{ac}\delta_{bd})\sin\left({{k_1\wedge k_4}\over{2}}\right)\sin\left({{k_2\wedge k_3}\over{2}}\right) \\
\left. \phantom{V^H_{abcd}(k_1,k_2,k_3,k_4)=}{}
+(\delta_{ad}\delta_{bc}-\delta_{ab}\delta_{cd})\sin\left({{k_3\wedge
k_1}\over{2}}\right)\sin\left({{k_2\wedge
k_4}\over{2}}\right)\right].
\end{gather*}
\end{itemize}

\subsection*{Acknowledgements}

It is a pleasure to thank the organisers of the XVIIth
International Colloquium on Integrable Systems and Quantum
Symmetries for their kind invitation. Most of the results
presented in this paper have been obtained from various
collaborations with E.~Cagnache, A.~de Goursac and T.~Masson.
Fruitful discussions with M.~Dubois-Violette and J.~Madore are
gratefully acknowledged.

\pdfbookmark[1]{References}{ref}
\LastPageEnding

\end{document}